\documentclass[acmsmall,screen,table]{acmart}

\AtBeginDocument{%
  }

\setcopyright{acmlicensed}
\copyrightyear{2025}
\acmYear{2025}
\acmDOI{XXXXXXX.XXXXXXX}

\acmISBN{978-1-4503-XXXX-X/18/06}
\usepackage[utf8]{inputenc}
\usepackage{amsmath}
\usepackage{enumerate}
\usepackage{threeparttable}
\usepackage{booktabs} 
\usepackage{graphicx}
\usepackage{calc}
\usepackage{xcolor}
\usepackage{color}
\definecolor{light-gray}{gray}{0.80}
\definecolor{grannysmithapple}{rgb}{0.66, 0.89, 0.63}
\definecolor{green(html/cssgreen)}{rgb}{0.0, 0.5, 0.0}
\definecolor{brightmaroon}{rgb}{0.76, 0.13, 0.28}

\usepackage{framed}
\usepackage{amsmath}

\newlength{\DepthReference}
\newlength{\HeightReference}
\newlength{\Width}%

\newcommand{\MyColorBox}[2][red]%
{%
    \settowidth{\Width}{#2}%
    \colorbox{#1}%
    {%
        \raisebox{-\DepthReference}%
        {%
                \parbox[b][\HeightReference+\DepthReference][c]{\Width}{\centering#2}%
        }%
    }%
}

\usepackage{pifont}
\usepackage{arydshln} 
\usepackage{listings}
\usepackage{lipsum}
\lstdefinestyle{interfaces}{
  float=tp,
  floatplacement=tbp,
  abovecaptionskip=0pt,
} 

\definecolor{codegreen}{rgb}{0,0.6,0}  
\lstdefinelanguage{diff}{  
  morecomment=[f][\color{blue}]{@@},     
  morecomment=[f][\color{red}]-,         
  morecomment=[f][\color{codegreen}]+,       
  morecomment=[f][\color{red}]{---}, 
  morecomment=[f][\color{codegreen}]{+++},
}

\usepackage{graphicx}

\usepackage{subcaption}
\usepackage{cleveref}

\usepackage{enumitem} 
\usepackage{booktabs}
\usepackage{multirow}
\usepackage{makecell}

\usepackage{pgfplots}

\usepackage{colortbl} 
\usepackage{rotating}

\usepackage{enumitem}

\usepackage{xspace}
\usepackage{xcolor}
\usepackage{url}
\usepackage{makecell} 
\usepackage{array}    
\usepackage{colortbl}
\usepackage{arydshln}
\usepackage[most]{tcolorbox}

\begin{document}

\title[\textsc{Fixseeker}]{\textsc{Fixseeker}: An Empirical Driven Graph-based Approach for Detecting Silent Vulnerability Fixes in Open Source Software}


\author{Yiran Cheng}
\affiliation{%
 \institution{Beijing Key Laboratory of IOT Information Security Technology, Institute of Information Engineering, CAS; School of Cyber Security, University of Chinese Academy of Sciences}
 \city{Beijing}
 \country{China}}
\email{chengyiran@iie.ac.cn}

\author{Ting Zhang}
\affiliation{%
 \institution{Singapore Management University}
 \city{Singapore}
 \country{Singapore}}
\email{tingzhang.2019@phdcs.smu.edu.sg}

\author{Lwin Khin Shar}
\affiliation{%
 \institution{Singapore Management University}
 \city{Singapore}
 \country{Singapore}}
\email{lkshar@smu.edu.sg}

\author{Zhe Lang}
\affiliation{%
 \institution{Beijing Key Laboratory of IOT Information Security Technology, Institute of Information Engineering, CAS; School of Cyber Security, University of Chinese Academy of Sciences}
 \city{Beijing}
 \country{China}}
\email{langzhe@iie.ac.cn}

\author{David Lo}
\affiliation{%
 \institution{Singapore Management University}
 \city{Singapore}
 \country{Singapore}}
\email{davidlo@smu.edu.sg}

\author{Shichao Lv}
\affiliation{%
 \institution{Beijing Key Laboratory of IOT Information Security Technology, Institute of Information Engineering, CAS; School of Cyber Security, University of Chinese Academy of Sciences}
 \city{Beijing}
 \country{China}}
\email{lvshichao@iie.ac.cn}

\author{Dongliang Fang}
\affiliation{%
 \institution{Beijing Key Laboratory of IOT Information Security Technology, Institute of Information Engineering, CAS; School of Cyber Security, University of Chinese Academy of Sciences}
 \city{Beijing}
 \country{China}}
\email{fangdongliang@iie.ac.cn}

\author{Zhiqiang Shi}
\affiliation{%
 \institution{Beijing Key Laboratory of IOT Information Security Technology, Institute of Information Engineering, CAS; School of Cyber Security, University of Chinese Academy of Sciences}
 \city{Beijing}
 \country{China}}
\email{shizhiqiang@iie.ac.cn}

\author{Limin Sun}
\affiliation{%
 \institution{Beijing Key Laboratory of IOT Information Security Technology, Institute of Information Engineering, CAS; School of Cyber Security, University of Chinese Academy of Sciences}
 \city{Beijing}
 \country{China}}
\email{sunlimin@iie.ac.cn}

\renewcommand{\shortauthors}{Cheng et al.}

\begin{abstract}
Open source software (OSS) vulnerabilities pose significant security risks to downstream applications. While vulnerability databases provide valuable information for mitigation, many security patches are released \emph{silently} in new commits of OSS repositories without explicit indications of their security impact. This makes it challenging for software maintainers and users to detect and address these vulnerability fixes. There are a few approaches for detecting vulnerability-fixing commits (VFCs) but most of these approaches leverage commit messages, which would miss \emph{silent} VFCs. On the other hand, there are some approaches for detecting silent VFCs based on code change patterns but they often fail to adequately characterize vulnerability fix patterns, thereby lacking effectiveness. 
For example, some approaches analyze each hunk in known VFCs, in isolation, to learn vulnerability fix patterns; but vulnerabiliy fixes are often associated with multiple hunks, in which cases correlations of code changes across those hunks are essential for characterizing the vulnerability fixes.    

To address these problems, we first conduct a large-scale empirical study on 11,900 VFCs across six programming languages, in which we found that over 70\% of VFCs involve multiple hunks with various types of correlations. Based on our findings, we propose \textsc{Fixseeker}, a graph-based approach that extracts the various correlations between code changes at the hunk level to detect silent vulnerability fixes. 
Our evaluation demonstrates that \textsc{Fixseeker} outperforms state-of-the-art approaches across multiple programming languages, achieving a high F1 score of 0.8404 on average in balanced datasets and consistently improving F1 score, AUC-ROC and AUC-PR scores by 32.40\%, 1.55\% and 8.24\% on imbalanced datasets. Our evaluation also indicates the generality of \textsc{Fixseeker} across different repository sizes and commit complexities.
\end{abstract}





\maketitle

\section{Introduction}
With the increasing adoption of open-source software (OSS) components, vulnerabilities embedded within them can propagate to a vast number of downstream applications~\cite{wu2023background}. 
One example is the critical OpenSSL vulnerability (CVE-2022-3602~\cite{CVE-2022-3602}) discovered in 2022, which affected numerous organizations and systems that rely on this library. This vulnerability in X.509 certificate verification could allow attackers to trigger a buffer overflow during certificate parsing, potentially leading to remote code execution.
Such incidents highlight how vulnerabilities in widely-used OSS can have far-reaching consequences across the software ecosystem. 

While vulnerability databases like the National Vulnerability Database (NVD) provide valuable information for mitigation, many security patches are released silently by making new commits in OSS repositories without explicit indications about vulnerability fixes and their security impacts. This practice, known as \emph{silent fixing}, follows coordinated vulnerability disclosure guidelines that recommend avoiding any reference to a security-related nature in commit messages~\cite{vulfixminer,householder2017cert,wang2020empirical, wang2019detecting}. For instance, a critical remote code execution vulnerability (CVE-2017-7645~\cite{CVE-2017-7645}) in the Linux kernel's NFS server implementation was silently fixed in March 2017, but not disclosed until September 2017, leaving users exposed to potential attacks during the six-month period.
The delay between patch availability and vulnerability disclosure creates a critical security challenge. Malicious actors can potentially exploit these vulnerabilities by analyzing public code changes before users know about them. 
This growing gap between patch availability and vulnerability awareness demonstrates the urgent need for automated tools to identify real-time security-relevant changes accurately.


To address these problems, automated approaches for identifying vulnerability-fixing commits (VFCs) have been proposed.
Many approaches~\cite{nguyen2022hermes, nguyen2022vulcurator, dunlap2024vfcfinder}  rely on textual analysis of commit messages or other metadata, containing security-related terms. While this can be effective for explicitly labeled security patches, it misses ``silent'' security fixes where developers intentionally omit security-related terms. Furthermore, many commit messages are often unclear, leading to the comprehension difficulty of the commit intention~\cite{ebert2019confusion, al2024detecting}. 
There are some approaches~\cite{vulfixminer, nguyen2023midas} that attempt to address the silent security fix problem by learning patterns solely from code changes to identify VFCs.
VulFixMiner~\cite{vulfixminer} extracts features from the diff between commits, while Midas~\cite{nguyen2023midas} employs multiple feature extractors at different code granularities.
These code-based approaches show promise but they face limitations in handling the structural complexity of vulnerability fixes. Many VFCs are composed of multiple hunks~\cite{wang2019detecting}, where a hunk represents a contiguous group of modified (i.e., added and removed) lines in a single file. Changes in these hunks of the VFC may be correlated in terms of fixing the vulnerability.
VulFixMiner and Midas both treat code changes as basic sequential structures, failing to consider the semantic relationships and dependencies between different hunks of code change, which may be crucial for machine learning-based approaches to adequately learn the vulnerability fix patterns.
Therefore, to effectively model such semantic relationships in VFC detection, we first need to understand their characteristics.
However, the challenge is that \textit{currently it is unclear what types of relationships exist between the different hunks of vfcs}.


\vspace{4px}
\noindent\textbf{Empirical Study.}
To address this challenge, we conduct an empirical study on a large-scale dataset containing 11,900 VFCs across six popular programming languages. Our analysis reveals that the majority of VFCs (over 70\% across all languages) involve multiple hunks. Through manual analysis, we observe that there are \emph{four} main types of correlations between hunks in multi-hunk VFCs: \emph{Caller-Callee Dependency}, \emph{Data Flow Dependency}, \emph{Control Dependency}, and \emph{Patten Replication}. These correlations exist in 93.03\% of the examined multi-hunk VFCs. 
Furthermore, we find that multi-hunk VFCs are more prevalent in fixing severe vulnerabilities, with 68.51\% of multi-hunk VFCs addressing ``critical'' and ``high'' security vulnerabilities, compared to 57.25\% for single-hunk VFCs.
These findings highlight the importance of considering inter-hunk correlations in VFC detection approaches.

\vspace{4px}
\noindent\textbf{Our Approach.}
Based on the findings from our empirical study, we proposed \textsc{Fixseeker}, a graph-based approach for detecting silent VFCs in OSS repositories.
At a high level, \textsc{Fixseeker} is a novel, synergistic combination of the use of code property graphs capturing the relationships among hunks, static analysis of code property graphs extracting features that characterize vulnerability fix patterns based on our empirical findings, and graph-based learning approach to effectively learn the features and classify VFCs.
More specifically, \textsc{Fixseeker} consists of \emph{five steps}.
First, given a commit, \textsc{Fixseeker} retrieves pre-commit and post-commit codes and preprocess both versions, then generates code property graphs (CPGs) for them. 
Second, it conducts static analysis in CPGs to extract the hunks' explicit correlation, containing the caller-callee dependency, data flow dependency, and control dependency. It uses code mapping with the Levenshtein algorithm to detect the implicit correlations (pattern replication). 
Third, it constructs Hunk Correlation Graphs (HCGs) for both pre-commit and post-commit states, then merges them into a unified CommitHCG.
Then, it embeds nodes with CodeBERT~\cite{feng2020codebert} and encodes edges as 4-D binary vectors to convert the CommitHCG into a numeric format.
Finally, CommitGNN uses Graph Convolutional Networks (GCNs) with an edge attention mechanism to learn the embedded CommitHCG. This model, trained with a weighted loss function to address the class imbalance, classifies whether a commit fixes a vulnerability. 

\vspace{4px}
\noindent\textbf{Evaluation.}
To evaluate \textsc{Fixseeker}, we use two benchmark datasets: a balanced dataset with a 1:1 ratio of VFCs to non-VFCs, and an imbalanced dataset with a 1:25 ratio of VFCs to non-VFCs. The balanced dataset allows us to assess the model's performance under ideal conditions, while the imbalanced dataset reflects the natural distribution in real-world scenarios where VFCs are rare. Our datasets cover four programming languages: C/C++, Java, Python, and PHP, with a total of 10,258 VFCs across 2,094 open-source projects.

To demonstrate the effectiveness of \textsc{Fixseeker}, we compare it with four state-of-the-art (SOTA) VFC detection approaches, namely PatchRNN~\cite{wang2021patchrnn}, VFFINDER~\cite{nguyen2023vffinder}, VulFixMiner~\cite{vulfixminer} and Midas~\cite{nguyen2023midas}, on both datasets. The results show that: i) \textsc{Fixseeker} outperforms other approaches in most metrics across different languages, achieving the highest F1-scores of 0.8577, 0.8447, 0.8319, and 0.8272 for four languages respectively on balanced datasets; ii) On imbalanced datasets, \textsc{Fixseeker} consistently demonstrates superior performance, 
improving the F1 by 32.4\%, AUC-ROC by 1.55\%, AUC-PR by 8.24\% on average compared to the best baseline methods. 
In terms of detecting the capability of different vulnerability types, we find that \textsc{Fixseeker} performs strongly in memory-related vulnerabilities, achieving recalls exceeding 0.75 for these types.
Moreover, to evaluate the efficiency of \textsc{Fixseeker}, we conduct an overhead analysis across six popular repositories of varying sizes. The results show that our CommitHCG generation process is efficient, achieving average generation time per commit ranging from 12.06 seconds to 30.92 seconds.

\noindent\textbf{Contribution.} This work makes the following contributions.
\begin{itemize}[leftmargin=1em]
    \item We conduct a large-scale empirical study to investigate the multi-hunk VFCs and the correlations within different hunks.

    \item Based on our empirical study, we propose a novel graph-based automated approach, named \textsc{Fixseeker}, to detect silent vulnerability fixes, supporting cross-language scenarios.

    \item We conduct extensive experiments to demonstrate the effectiveness, generality and practical usefulness of \textsc{Fixseeker}.
\end{itemize}

The rest of the paper is organized as follows. Section 2 introduces the preliminaries and a motivating example. 
Section 3 presents our empirical study on multi-hunk VFCs. 
Motivated by the findings of empirical study, Section 4 details our proposed approach \textsc{Fixseeker}. 
Section 5 presents the experimental setup and evaluation results. 
Section 6 discusses our approach's practicality, limitations, and future work. Section 7 reviews related work and Section 8 concludes the paper.
\section{Preliminaries and Motivation}

\subsection{Definition}
\noindent\textbf{VFCs and non-VFCs.}
The commits of open-source software record the changes between two different source code versions. 
In our work, we define a "commit" as a single-purpose Git commit. 
We consider a commit as a "vulnerability-fixing commit" (VFC) if it fixes a vulnerability belonging to any Common Weakness Enumeration Specification (CWE) type.
Conversely, we define a "non-vulnerability-fixing commit" (non-VFC) as any commit that does not address a security vulnerability, including those that fix functional bugs, add new features, or perform code refactoring.

\noindent\textbf{Commit Hunk.}
A commit hunk represents a contiguous group of changed lines within a single file in a code commit. Each hunk contains a set of lines that were either added or removed. Formally, a hunk can be defined as a tuple $H = (-S, O, +S', N)$, where $S$ ($S'$) is the starting line number in the original (new) file, $O$ ($N$) is the number of modified lines in the original (new) file. 
As shown in Figure~\ref{fig:example}, the VFC has three hunks in the modified file \texttt{iwl-agn-sta.c}, the corresponding value of $(-S, O, +S', N)$ in the first hunk is (-35,9,+35,12).
Hunks are fundamental units in understanding and analyzing code changes, as they provide a granular view of the modifications made in a commit. In the context of VFCs, analyzing hunks and their relationships can offer insights into the nature and extent of security-related code changes.

\noindent\textbf{Code Property Graph (CPG).} 
A CPG is a unified graph representation that combines multiple program analysis graph types to capture different aspects of code semantics. Formally, a CPG can be defined as $G = (V, E)$, where $V$ is a set of nodes representing program elements (e.g., functions, variables, operations), $E$ is a set of directed edges connecting these nodes. CPG integrates the nodes and edges from three fundamental program analysis graphs: Abstract Syntax Tree (AST) graph representing the syntactic structure, Control Flow Graph (CFG) capturing control dependencies, and Data Flow Graph (DFG) representing data dependencies. 

\subsection{Motivating Example}

\begin{figure}[t]
    \centering
    \includegraphics[width=1.0\linewidth]{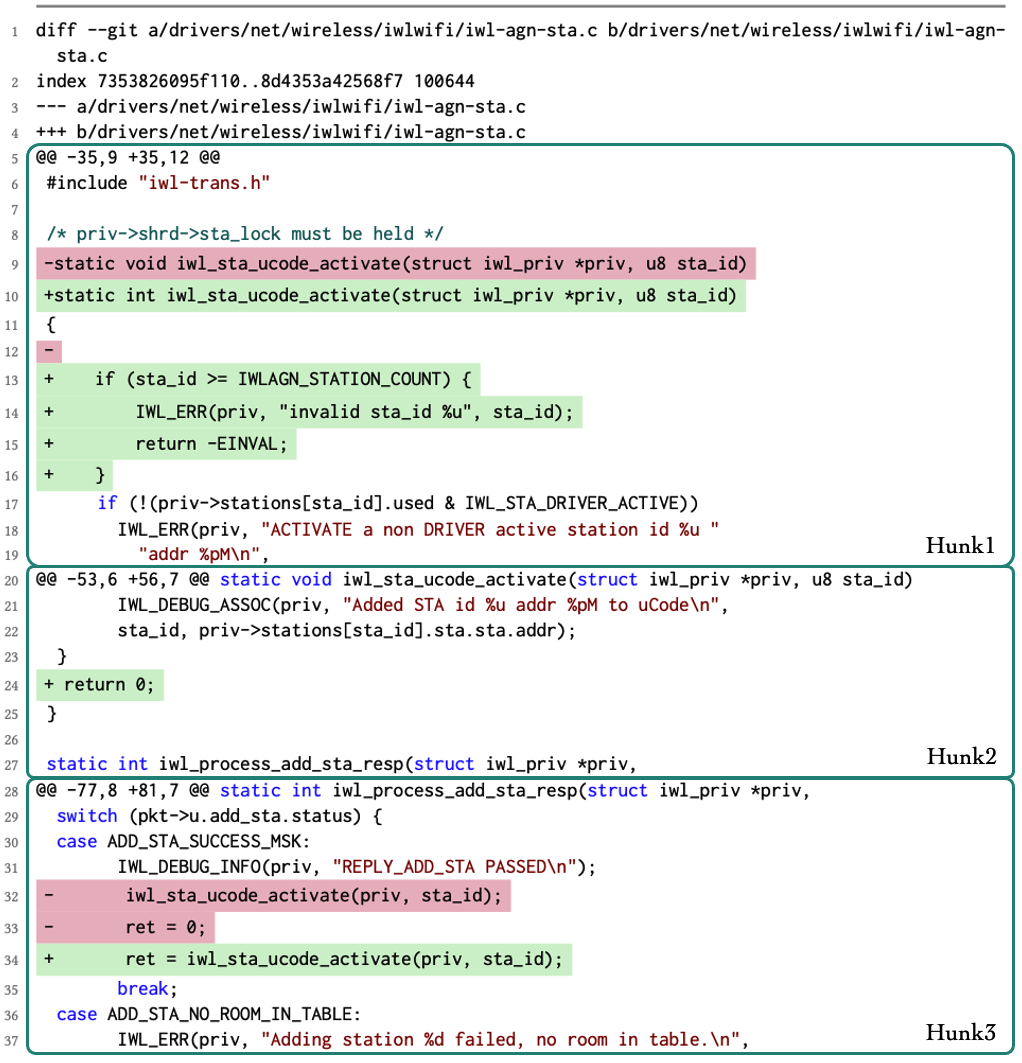}
    \caption{Motivating Example of CVE-2012-6712.}
    \label{fig:example}
\end{figure}

To motivate our work, we present a vulnerability fix in the Linux kernel's wireless driver that demonstrates how multiple hunks collaborate to address a security vulnerability, highlighting the importance of understanding hunk relationships in vulnerability fixes. 
In this case, shown in Figure~\ref{fig:example}, three semantically related hunks collectively fix a potential out-of-bounds vulnerability in the \texttt{iwl\_sta\_ucode\_activate} function. The first hunk (Lines 9-16) modifies the function signature from \texttt{void} to \texttt{int} and adds a bounds check \texttt{if (sta\_id >= IWLAGN\_STATION\_COUNT)} to validate the station ID parameter. The second hunk (Line 24) adds a \texttt{return 0} statement to handle the success case properly. 
These two hunks together transform the function to enforce proper bounds checking and return status indication. 
The third hunk (Lines 32-34) adapts the caller side by replacing the direct function call with proper error handling of the newly added return value. These hunks exhibit a caller-callee dependency: modifying the callee function's signature and behavior (first two hunks) necessitates corresponding changes in the caller's code (third hunk).

Current approaches like Midas~\cite{nguyen2023midas}, although extracting features at multiple granularities including commit, file, hunk, and line levels, focus primarily on analyzing each hunk independently when designing their feature extractors. 
When applied to this case, it processes the addition of bounds checking (Lines 13-15) and the modification of return value handling (Lines 32-34) as separate features, failing to capture their inherent caller-callee relationship that is crucial for understanding how the complete vulnerability fix works through coordinated error handling modifications across multiple functions. If these semantic relationships between hunks could be explicitly modeled and learned, the model would be better equipped to recognize such coordinated security fixes.

This example reveals a critical challenge in VFC detection: vulnerability fixes often involve multiple coordinated changes (hunks) that are semantically related but syntactically dispersed. 
By modeling these inter-hunk relationships in a graph structure,  we can learn richer embeddings from graph representation that incorporate both the content of individual hunks and their relationships with other hunks, leading to more accurate detection of vulnerability fixes. 
However, to effectively model these semantic relationships in VFC detection, we first need to understand what types of relationships exist between different hunks of VFCs.
This observation motivates us to conduct a large-scale empirical study on the relationships between different hunks in VFCs, aiming to understand the patterns and types of inter-hunk correlations contributing to security fixes.

\section{An Empirical Study}
\label{sec:empirical_study}
We design an empirical study to understand the characteristics of VFC hunks for OSS vulnerabilities in databases. We aim to answer the following research questions.

\begin{itemize}[leftmargin=1em]
\item \textbf{RQ1: Multi-hunk Prevalence: } What proportion of VFCs contain multiple code hunks?  (Sec.~\ref{sec:coverage})
\item \textbf{RQ2: Hunk Correlation Analysis: } What types of correlations exist between different hunks in multi-hunk VFCs? (Sec.~\ref{sec:hunk_relation})
\item \textbf{RQ3: Severity and Vulnerability Type Analysis: } How does vulnerability severity relate to multi-hunk VFCs? (Sec.~\ref{sec:severity_type}) 
\end{itemize}

\subsection{Data Preparation}
We selected six popular PLs as our analysis subjects: C/C++, Java, Python, PHP, JavaScript, and Golang. 
We extracted all CVEs related to the six languages disclosed by September 26, 2024 from NVD~\cite{NVD}.
For each vulnerability entry in NVD, we analyze its reference section, which contains links to various resources. From these references, we identify GitHub commit URLs by filtering for links that contain ``github.com'' and ``commit''. These commits typically represent the fixes for the reported vulnerabilities.
A commit may contain code changes in various types of files, but we focus only on .c, .cpp, .java, .py, .php, .js, and .go files. Therefore, we filtered the dataset by removing the commits that only contained modification files unrelated to the target languages.
Finally, we collected 5,168 commits in C/C++, 786 commits in Java, 1,147 commits in Python, 3,157 commits in PHP, 1,039 commits in JavaScript, and 603 commits in Golang, across 2,832 open-source projects. As shown in Table~\ref{tab:count}, C/C++ projects contributed the most VFCs. 
We observed that some commits include modifications to test files. However, since the modifications of test files are typically intended to verify the effectiveness of vulnerability fixes rather than containing the actual vulnerability fix code, we ignore them in our analysis.

\begin{table}[h]\small
    \centering
    \setlength{\tabcolsep}{6mm}
    \caption{The Dataset Distribution across Different Program Languages} 
    \vspace{-2mm}
    \begin{tabular}{l|cccc}
        \Xhline{0.8pt}
        \textbf{PL} & \textbf{\#VFC} & \textbf{\#File} & \textbf{\#Hunk} & \textbf{\#Project} \\
        \hline \hline
        C/C++ & 5,168 & 13,148 & 35,723 & 720 \\
        Java & 786 & 4,281 & 9,817 & 260\\
        Python & 1,147 & 4,990 & 12,163 & 385\\
        PHP & 3,157 & 27,907 & 60,162 & 729 \\
        JavaScript & 1,039 & 8,267 & 16,857 & 527 \\
        Golang & 603 & 3,695 & 9,756 & 211\\
        \hline
        Overall & 11,900 & 62,288 & 144,478 & 2,832 \\
        \Xhline{0.8pt}
    \end{tabular}
    \label{tab:count}
\end{table}
\vspace{-2mm}

\subsection{Multi-hunk Prevalence (RQ1)}
\label{sec:coverage}
\begin{figure}[h]
    \centering
    \includegraphics[width=0.82\linewidth]{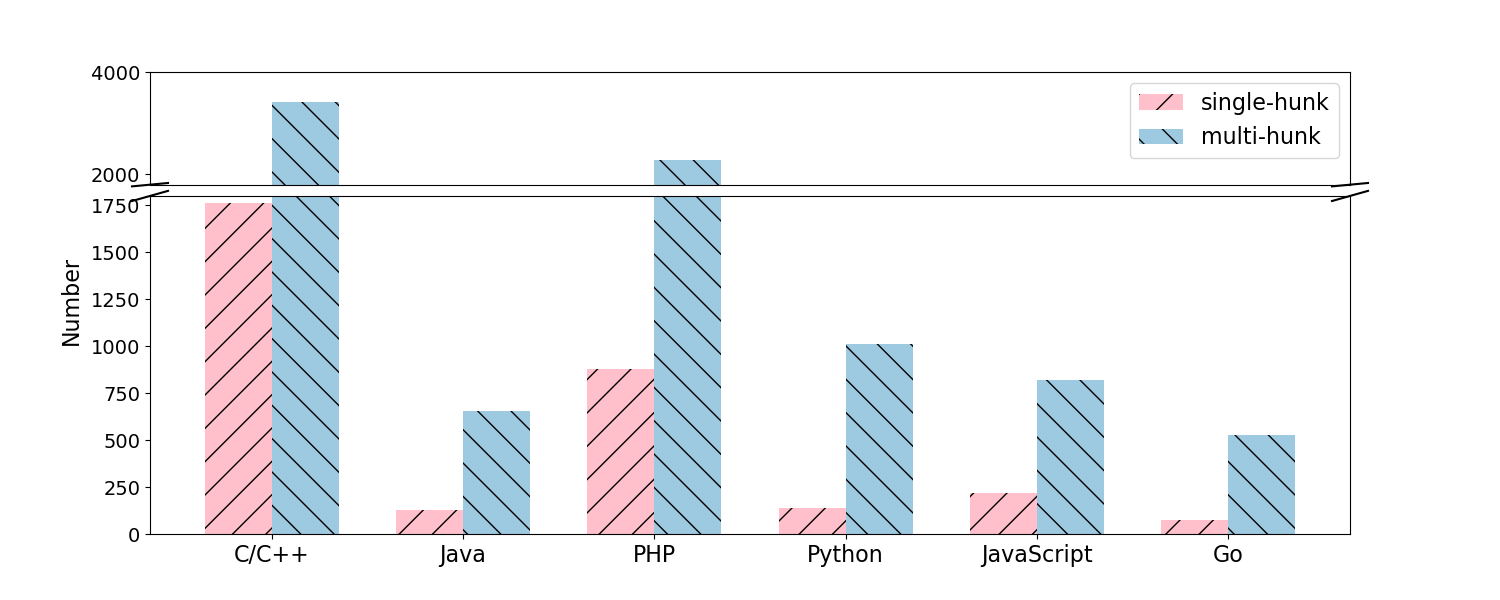}
    \vspace{-2mm}
    \caption{The Number of Multi-hunk and Single-hunk VFCs across Different Program Languages}
    \label{fig:multi_and_single}
\end{figure}

Table~\ref{tab:count} shows some statistics of the dataset. C/C++ has the highest number of VFCs at 5,168, accounting for 43.4\% of the total. This is followed by PHP (3,157, 26.5\%) and Python (1,147, 9.6\%). 
PHP contains the most modified files and hunks. On average, one file contains 2.32 code change hunks.
In terms of project coverage, C/C++ involves 720 projects, significantly higher than other languages.
Examining the complexity of modifications, we categorized VFCs into single-hunk and multi-hunk changes. 
Figure~\ref{fig:multi_and_single} illustrates that across all languages, the number of multi-hunk modifications significantly exceeds single-hunk modifications, indicating that most vulnerability fixes involve multiple code changes. This is particularly evident in Python and Golang, where the number of multi-hunk modifications far exceeds single-hunk modifications over 7 times.
\begin{center}
\fcolorbox{black}{gray!10}{\parbox{0.98\linewidth}{\textbf{Finding\#1}: Across all studied PLs, multi-hunk fixes significantly outnumber single-hunk fixes. This is particularly evident in Python and Golang, where multi-hunk fixes exceed single-hunk fixes by over 7 times.}}
\end{center}

\vspace{-2mm}

\subsection{Hunk Correlation Analysis (RQ2)}
\label{sec:hunk_relation}
\begin{figure}[tp]
    \centering
    \includegraphics[width=1.0\linewidth]{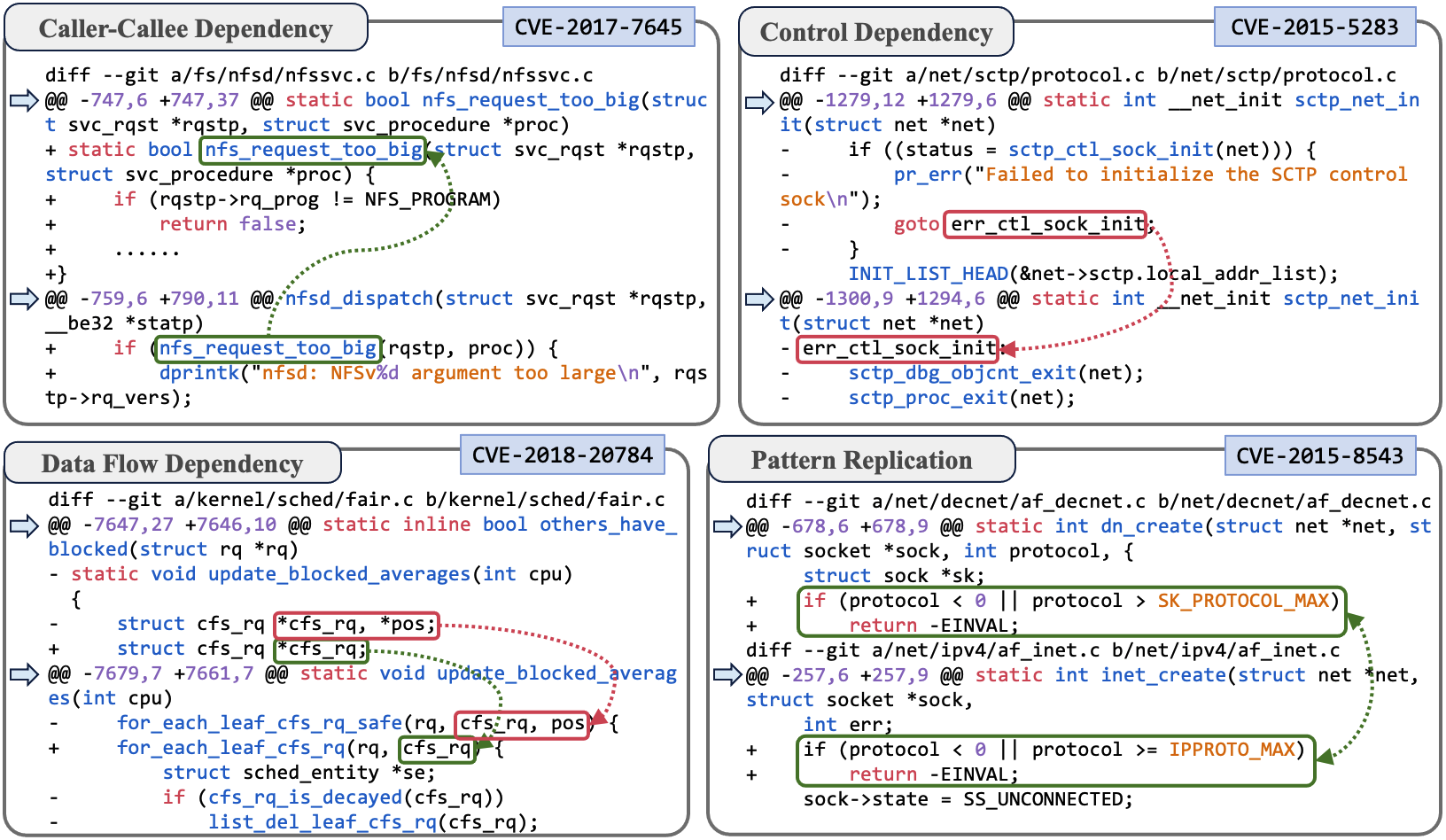}
    \caption{Examples for Different Hunk Correlations}
    \label{fig:hunk_relation}
    \vspace{-2mm}
\end{figure}

To investigate this RQ, we conduct a manual analysis of sample VFCs.
We considered a 95\% confidence level with a 5\% margin of error, resulting in a sample size of 373 VFCs out of the 11,900 total VFCs in the dataset.
We randomly selected 373 
multi-hunk VFCs from popular open-source project Linux~\cite{Linux}, FFmpeg~\cite{FFmpeg} and ImageMagick~\cite{imagemagick} in the C/C++ VFC dataset to manually analyze the correlations between hunks. 
We observed that the hunks in 93.03\% commits have correlations in terms of Caller-Callee Dependency, Data Flow Dependency, Control Dependency and Pattern Replication, as shown in Table~\ref{tab:correlation}. Figure~\ref{fig:hunk_relation} illustrates the above four types of correlation, we use arrows to indicate the starting points of different hunks. The four types of correlations are explained below. 

1) Caller-Callee Dependency: The hunk in function $A$ has code changes (i.g., function addition and modification), and the call-related hunk of $A$'s caller also undergoes code changes. In the VFC of CVE-2017-7645~\cite{CVE-2017-7645} in Figure~\ref{fig:hunk_relation}, the second hunk calls the function \texttt{nfs\_request\_too\_big} in the first hunk. This type of correlation accounts for 35.62\% of commits in Linux, 21.15\% in FFmpeg, and 10.96\% in ImageMagick.

2) Data Flow Dependency: The code in hunk $B$ uses the data defined or assigned in hunk $A$. 
In the VFC of CVE-2018-20784~\cite{CVE-2018-20784} in Figure~\ref{fig:hunk_relation}, the second hunk uses the pointer variables \texttt{cfs\_rq} and \texttt{pos} from the first hunk.
This type of correlation is particularly prominent in FFmpeg, accounting for 52.46\% of commits, while it represents 35.16\% in Linux and 43.84\% in ImageMagick.

3) Control Dependency: The code in hunk $B$ have a control dependency on hunk $A$. When the code in hunk $A$ changes, hunk $B$ changes accordingly. 
In the VFC of CVE-2015-5283~\cite{CVE-2015-5283} in Figure~\ref{fig:hunk_relation}, the second hunk has a control dependency on the first hunk because of the statement \texttt{goto err\_ctl\_sock\_init}.
This type of correlation is most prevalent in FFmpeg at 54.10\%, followed by 41.10\% in ImageMagick and 27.85\% in Linux.

4) Pattern Replication: The code in hunk $A$ and hunk $B$ have highly similar code change patterns. 
In the VFC of CVE-2015-8543~\cite{CVE-2015-8543} in Figure~\ref{fig:hunk_relation}, the two hunks both add two conditional checks for variable \texttt{protocol}.
This type of correlation is most common in ImageMagick, accounting for 52.05\% of commits, while it represents 42.01\% in Linux and 22.95\% in FFmpeg.

The first three are explicit correlations, and the last one is an implicit correlation. 
We observed that many commits contain multiple types of hunk correlations (correlation ``Hybrid'' in Table~\ref{tab:correlation}). The proportion of such commits in the three OSS projects is 37.9\%, 37.70\%, and 34.25\%, respectively.
\begin{center}
\fcolorbox{black}{gray!10}{\parbox{0.98\linewidth}{\textbf{Finding\#2}: Correlations exist in 93.03\% of multi-hunk VFCs. We identified four main correlation types: caller-callee Dependency, data flow dependency, control dependency, and pattern replication. Many commits (around 35\%) contain multiple correlation types.}}
\end{center}

\begin{table}[tp]\small
    \centering
    \setlength{\tabcolsep}{4mm}
    \caption{The Distribution of Different Hunk Correlations in Open-source Software}
    \vspace{-2mm}
    \begin{tabular}{c|l|c|c|c}
        \Xhline{0.8pt}
        \multirow{2}*{\textbf{Type}} & \multirow{2}*{\textbf{Correlation}} & \multicolumn{3}{c}{\textbf{Open-source Project}}\\
        \cline{3-5}
        ~ & ~ & Linux & FFmpeg & ImageMagick\\ [0.2ex]
        \hline
        \hline
        \multirow{3}*{Explicit} & Caller-Callee Dependency & \cellcolor{lightgray}35.62\% & 21.15\% & 10.96\% \\
        \cline{2-5}
        ~ & Data Flow Dependency & 35.16\% & \cellcolor{lightgray}52.46\% & 43.84\%\\
        \cline{2-5}
        ~ & Control Dependency & 27.85\% & \cellcolor{lightgray}54.10\% & 41.10\%\\
        \hline
        Implicit & Pattern Replication & 42.01\% & 22.95\% & \cellcolor{lightgray}52.05\%\\
        \hline
        - & Hybrid & 37.90\% & 37.70\% & 34.25\% \\
        \Xhline{0.8pt}
    \end{tabular}
    
    \label{tab:correlation}
    \vspace{-1mm}
\end{table}

\subsection{Severity Analysis (RQ3)}
\label{sec:severity_type}
We collected the severity of the vulnerabilities from NVD corresponding to the VFCs. NVD uses Common Vulnerability Scoring System (CVSS) to rate vulnerabilities on a scale of 0 to 10, the severity levels are Low (0-3.9), Medium (4.0-6.9), High (7.0-8.9) and Critical (9.0-10.0). Specifically, 5,962 (68.51\%) multi-hunk VFCs fix the vulnerability with severity tags ``Critical'' and ``High'', while for single-hunk VFCs, the number is 1,831 (57.25\%). This suggests that multi-hunk VFCs are more inclined to fix more severe vulnerabilities.

\begin{center}
\fcolorbox{black}{gray!10}{\parbox{0.98\linewidth}{\textbf{Finding\#3}: 68.51\% of multi-hunk VFCs fix "Critical" and "High" severity vulnerabilities, compared to 57.25\% for single-hunk VFCs. }}
\end{center}

\section{Our Approach}
In this section, we propose an automated approach, \textsc{Fixseeker}, for detecting VFCs in OSS repositories, without relying on commit messages. The motivation of \textsc{Fixseeker} is based on the findings from our empirical study that VFCs often address the vulnerability by changing multiple associated code hunks, exhibiting four types of correlation patterns. We leverage code graphs as they characterize dependencies among code changes and hunks. We employ graph neural networks as they can naturally learn complex patterns in graphs. 
In the following, we first provide an overview of our approach. We then explain the steps involved in detail, using a running example (Figure~\ref{fig:running_example}).

\begin{figure}[h]
    \centering
    \includegraphics[width=0.98\linewidth]{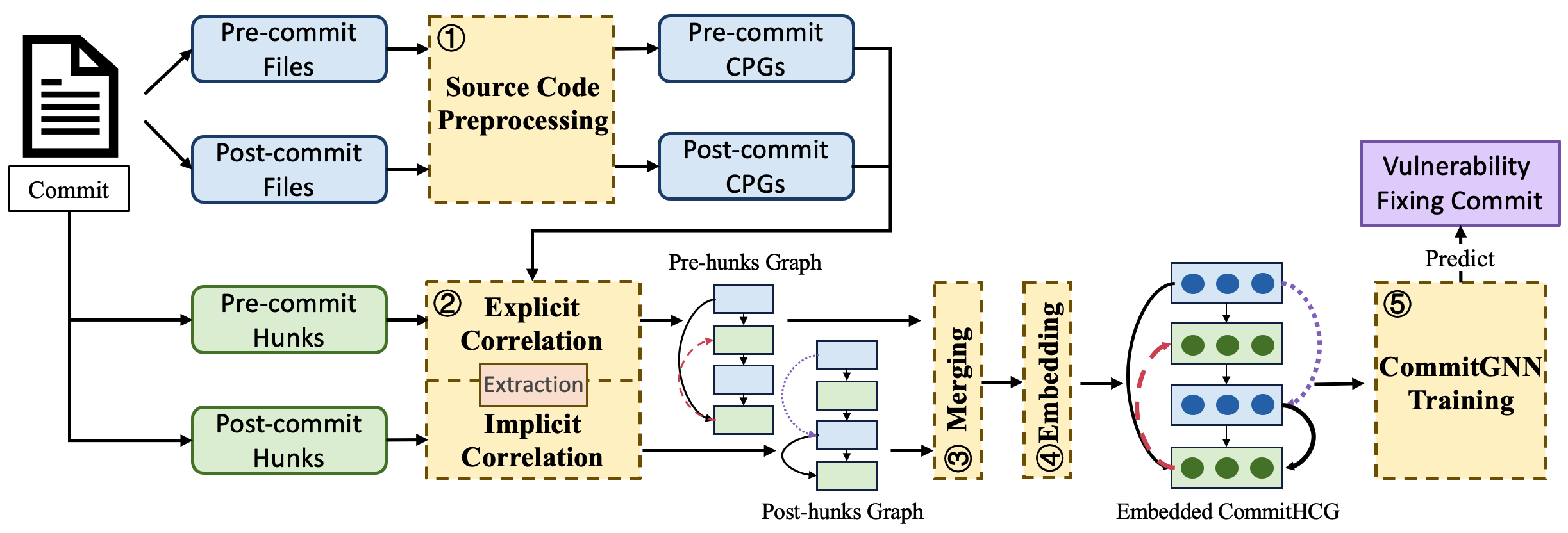}
    \caption{The Workflow of \textsc{Fixseeker}}
    \label{fig:overview}
\end{figure}

Figure~\ref{fig:overview} presents an overview of \textsc{Fixseeker}. 
\textsc{Fixseeker} takes a commit as its input and outputs whether the commit is for vulnerability-fixing or not.
\textsc{Fixseeker} works in five steps. 
1) First, it retrieves pre-commit and post-commit code via git reset, preprocesses functions, and generates Code Property Graphs (CPGs) for both versions. 
2) It then proceeds to extract the hunk correlations, conducting the static analysis in CPGs to analyze explicit correlations while using code mapping to identify implicit correlations, creating Hunk Correlation Graphs (HCGs). 
3) The Merging step combines pre-commit and post-commit HCGs into a unified CommitHCG. 
4) Then, it embeds nodes with CodeBERT and encodes edges as 4-D binary vectors. 
5) Finally, we propose a Graph Neural Network (GNN) model CommitGNN, which processes the embedded CommitHCG through GCN layers with edge attention mechanisms and a weighted loss function, and then predicts whether the corresponding commit fixes a vulnerability.

\subsection{Hunk Correlation Graph Construction}
A commit is composed of code changes in a group of files. A hunk represents a set of sequentially concatenating line(s) of code changes within a file. We decompose the diff code of commit $C$ into a set of hunks: $C = \{H_1, H_2, ..., H_n\}$.
Since a hunk contains both pre-commit and post-commit code, we analyze the correlation of hunk code in pre-commit and post-commit states separately, denoted as $H_{pre}$ and $H_{post}$, which are assigned the same $id$.

\subsubsection{Source Code Preprocessing (Step 1)}
Given a commit ID, we can use git reset to roll back and obtain the source code before and after the commit. By diffing the source code before and after the commit, we can identify the changed files, denoted as $F_{pre}$ and $F_{post}$.
We first identify all function and their respective scopes (defined by start and end line numbers) in both $F_{pre}$ and $F_{post}$ using Joern Parser~\cite{Joern}. The diff file provides range information for each code change hunk (i.g., Line 2 of CVE-2017-7645's VFC in Figure~\ref{fig:hunk_relation}), enabling us to locate the corresponding functions of hunks, denoted as $Fc_{pre}$ and $Fc_{post}$ via scope comparison.
To focus more specifically on the functions affected by code changes, we remove patch-irrelated functions in the files.  
Subsequently, we utilize the Joern parser again to generate Code Property Graphs (CPGs) for $F_{pre}$ and $F_{post}$ separately, yielding $CPG_{pre}$ and $CPG_{post}$ respectively.
Note that if the VFC contains only a single hunk, we can skip the hunk correlation extraction (Step 2) and merging process (Step 3) since there are no inter-hunk relationships to analyze. 

\subsubsection{Multi-hunk Correlation Extraction (Step 2)}
As we mentioned in Section~\ref{sec:hunk_relation}, the correlation in different hunks can be categorized into explicit and implicit correlations. Explicit correlation includes caller-callee, data flow, and control dependencies between hunks. Implicit correlation refers to hunk diffs that have similar code change patterns (pattern replication). 

1) \textit{Caller-callee dependency extraction.} 
Given a CPG, we traverse the functions in the CPG to obtain the caller and callee for each function, forming a function graph with two sets: $(N, E)$. $N$ is a set of nodes represented by function $ids$, where $id$ is the unique identifier assigned to each function by the CPG. $E$ is a set of directed edges represented by 2-tuples $(id1, id2)$, where $id1$ is the caller and $id2$ is the callee. If there is a call operation to the callee within the hunks of the caller, we add an edge between these two hunks, represented as $(H_{id1}, H_{id2}, type)$, $type\in\{CALL\}$.

2) \textit{Data flow and control dependency extraction.}
CPG consists of nodes and edges. Nodes are represented by $id$, $code$, and $line number$. The $code$ refers to a statement component in the source code, while the $line number$ indicates the line where this statement is located. Edges represent relations between the nodes, represented as $(id1, id2, type)$. There are three types of edges: Abstract Syntax Tree (AST), Data Dependency (DD), and Control Dependency (CD). We only consider relation DD and CD, since AST relations exist only among multiple components within a single statement, which is irrelevant to correlations between hunks.
We employ scope comparison to determine whether $ids$ with relations belong to different hunks, if so, we add an edge between these two hunks, represented as $(H_{id1}, H_{id2}, type)$, $tpye\in\{DD,CD\}$.

3) \textit{Pattern replication extraction.}
To determine whether similar modification patterns have been applied between hunks, we use a code mapping approach based on the Levenshtein distance algorithm. The Levenshtein distance measures the minimum number of single-character edits (insertions, deletions, or substitutions) required to change one string into another.
We first tokenize the hunk code snippets and represent each hunk as a sequence of tokens. For each pair of hunks $(H_{id1}, H_{id2})$, we calculate the normalized Levenshtein distance:
$NLD(H_{id1}, H_{id2})= 1 - (H_{id1}, H_{id2}) / max(len(H_{id1}), len(H_{id2})))$,
where $LD(H_{id1}, H_{id2})$ is the Levenshtein distance between the token sequences of $H_{id1}$ and $H_{id2}$, and $len(H)$ is the number of tokens in hunk $H$.
We set a similarity threshold $\theta = 0.8$ used in previous works~\cite{karnalim2019source}. If $NLD(H_{id1}, H_{id2}) > \theta$, we consider the hunks to have a similar modification pattern and add an edge between these two hunks, represented as $(H_{id1}, H_{id2}, type)$, where $tpye\in\{SIM\}$.

\subsubsection{Merging into CommitHCG (Step 3)} 
For each pair of pre-commit and post-commit hunks, we
merge the corresponding two HCGs into a unified graph structure called CommitHCG. 
We first merge the $H_{pre}$ and $H_{post}$ with the same $id$ as a unified hunk node. 
Then, we transform the source and end nodes of pre-commit and post-commit HCG edges into the unified hunk node.
In this way, we obtain a CommitHCG depicted by two sets $(N', E')$, where $N'$ is the set of unified hunk nodes and $E' = E_{pre} \cup E_{post}$, and $E'$ is represented with 3-tuples $(H_{id1}, H_{id2}, type)$, $type\in\{CALL,DD,CD,SIM\}$.

\subsubsection{Running Example}
\begin{figure}
    \centering
    \includegraphics[width=0.9\linewidth]{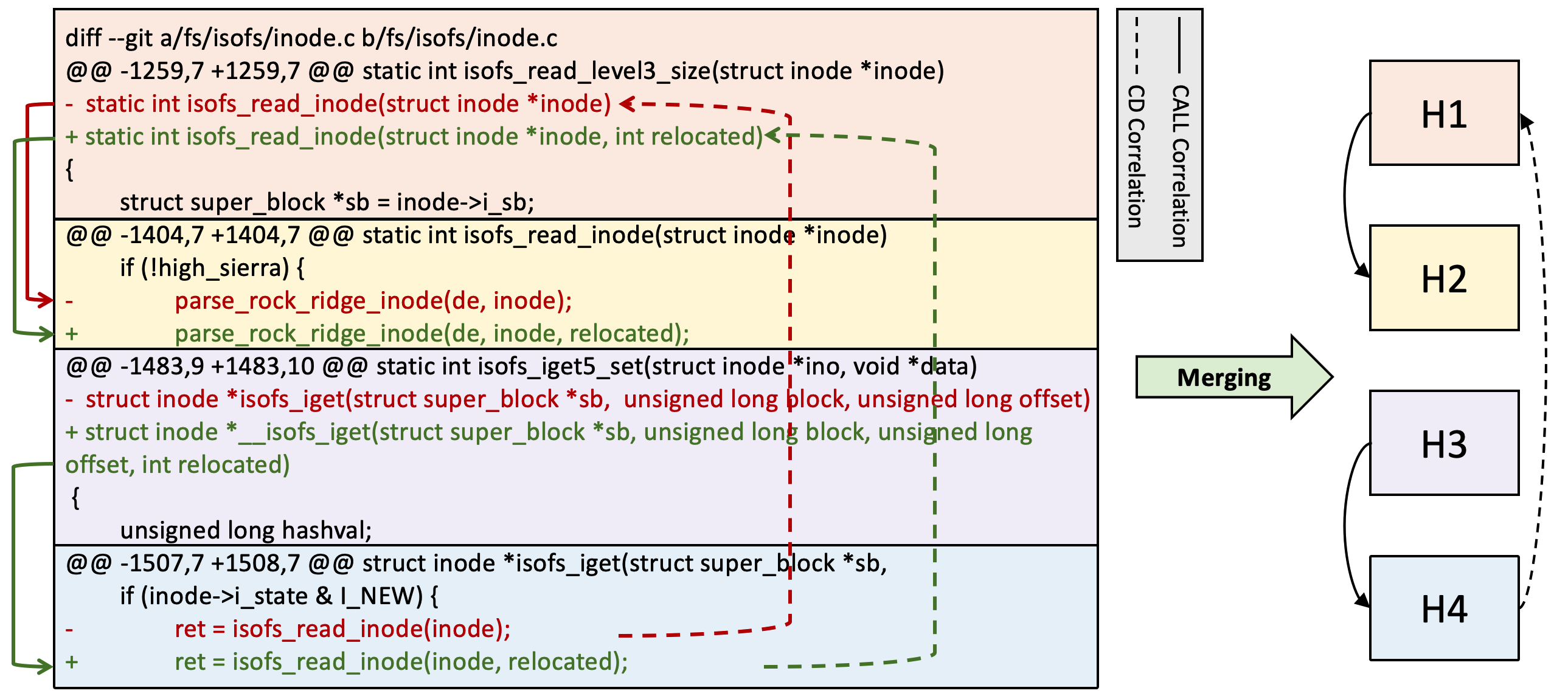}
    \caption{A Running Example of Hunk Correlation Graph Construction}
    \label{fig:running_example}
\end{figure}

We use the commit \texttt{410dd3} in Linux Kernel as the running example. Figure~\ref{fig:running_example} shows the code diff of the commit. 
In step 1, \textsc{Fixseeker} identifies the modified files (\texttt{fs/isofs/inode.c}) through the ``diff --git [a/file] [b/file]'' markers. Then it separates the four modified hunks within each file using the ``@@[-s,o,+s',n]@@'' markers. \textsc{Fixseeker} obtains the $F_{pre}$ and $F_{post}$ of the file using \texttt{git reset} to roll back the changes in the repository. The Joern parser is used to determine the scope of all functions in both versions.
By comparing the hunk ranges (s,o) and (s',n) with all the function scopes, \textsc{Fixseeker} locates the changed function \texttt{isof\_read\_inode} with one parameter and \texttt{isof\_iget} in $F_{pre}$, \texttt{isof\_read\_inode} with two parameters and \texttt{\_\_isof\_iget} in $F_{post}$. It then removes the other unchanged (patch-unrelated) functions from $F_{pre}$ and $F_{post}$.
Finally, \textsc{Fixseeker} generates the corresponding CPGs for the refined versions: $CPG_{pre}$ and $CPG_{post}$.

In step 2, \textsc{Fixseeker} extracts the correlations between hunks. In the caller-callee dependency extraction, it identifies function calls to \texttt{isof\_read\_inode} in \texttt{isof\_iget} of $F_{pre}$ and \texttt{\_\_isof\_iget} of $F_{post}$, creating \textit{CALL} edges between these hunks. For data flow dependency extraction, \textsc{Fixseeker} analyzes the CPGs and finds that within the \texttt{isof\_read\_inode} function, it finds \textit{DD} edges between the first hunk and the second hunk. In \texttt{\_\_isof\_iget} function, it creates \textit{DD} edges between the third hunk and the fourth hunk because of the dependency of variable \texttt{relocated}.
For pattern replication extraction, it tokenizes the code changes in each hunk and calculates the normalized Levenshtein distance between them but does not find similar code changes.

In step 3, \textsc{Fixseeker} merges the pre-commit and post-commit HCGs into a unified CommitHCG. The resulting graph contains four nodes representing the modified hunks, 
with three edges representing the identified correlations (\textit{CALL}, \textit{DD}).


\subsection{Graph Learning for Detecting Vulnerability Fixing Commit}
After constructing the CommitHCG, we employ graph learning techniques to detect VFCs. This process involves three main steps: graph embedding, where we transform node and edge information into numerical vectors; CommitGNN model design, which leverages Graph Neural Network (GNN) with edge attention mechanism to capture the complex relationships within the CommitHCG; and model training, where we address class imbalance issues and optimize the model's performance. 

\subsubsection{Graph Embedding (Step 4)}
To input CommitHCGs into a GNN-based model, the node and edge attributes in the graphs should be embedded as numeric vectors. Node attributes represent the Hunk code snippet in each node, while edge attributes represent the correlations between nodes.

1) \textit{Node embedding.}
Graph node is a pair of pre-commit (removed) and post-commit (added) codes within a hunk code snippet.
\textsc{Fixseeker} fine-tunes CodeBERT~\cite{feng2020codebert} as code embedding models for representing hunk code snippets in the graph nodes.
In the representation, we consider the joint input format of CodeBERT as follows:
$$[CLS]<-code>[SEP]<+code>[EOS]$$
where [CLS] token is placed at the beginning of the input sequence and is used to capture a representation of the entire input for classification tasks.
[SEP] token separates different segments (removed and added parts) within the input.
[EOS] marks the end of the input sequence.
<-code> represents removed codes, where each statement begins with the `-' symbol. <+code> represents added codes, where each statement begins with the `+' symbol.

2) \textit{Edge embedding.}
Edge embedding is used to reflect the relations between two nodes. The edge types in CommitHCGs involve four types of relations: CALL, CD, DD and SIM. Since the edges of CALL, CD and DD types are directed, we design the SIM type edges as bidirectional. Additionally, multiple edges may exist between two nodes. Therefore, the edge embedding is designed as a 4-dimensional binary vector, with each of the 4 bits indicating whether there exists any CALL, CD, DD, or SIM edge between the current two nodes, respectively. If two hunk nodes have CD and SIM relations, the edge embedding vector will be (0, 1, 0, 1). 

\subsubsection{CommitGNN Model (Step 5)}
\begin{figure}[h]
    \centering
    \includegraphics[width=0.75\linewidth]{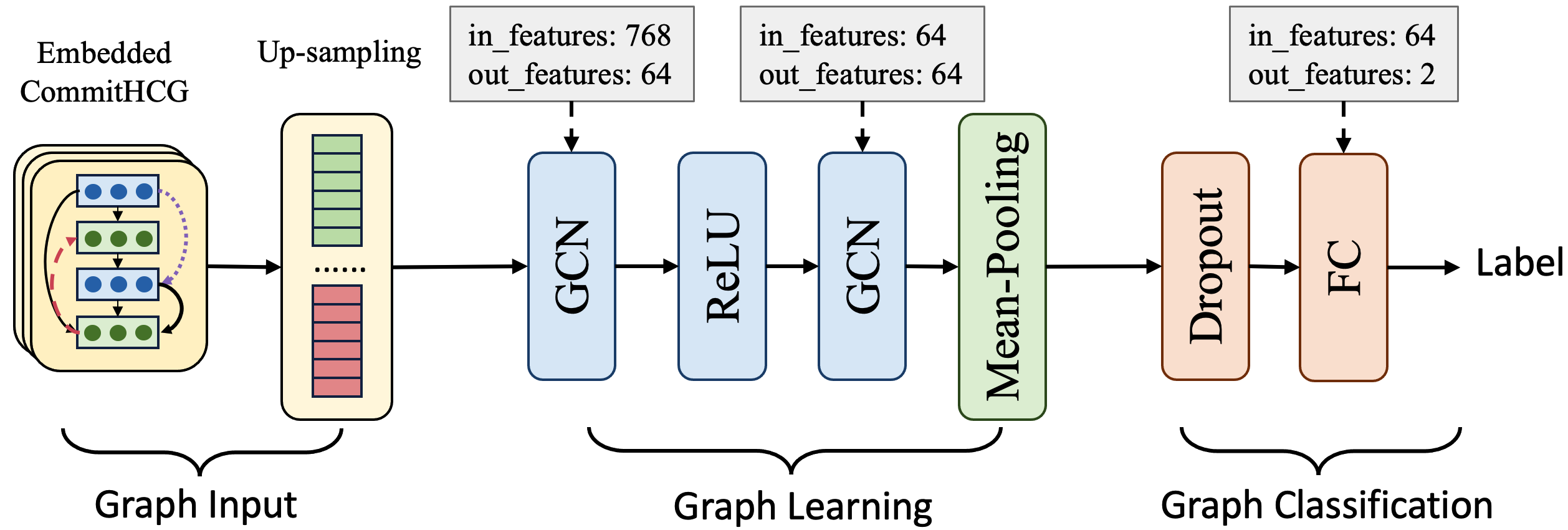}
    \caption{The Design of CommitGNN}
    \label{fig:model}
\end{figure}

Unlike conventional models focusing on matrix data, GNNs can directly capture complex node relations and dependencies in graph-structured data. 
In this section, we introduce our CommitGNN model to address the VFC detection problems. As illustrated in Figure~\ref{fig:model}, our model consists of three parts: graph input, graph learning, and graph classification. 

1) \textit{Graph input.}
The input graph to our model is CommitHCG ($\mathcal{G}$), consisting of nodes $\mathcal{V}=\{H_1,H_2,...,H_n\}$ and edges $\mathcal{E}$. Node features are represented by a matrix with dimensions ($|\mathcal{V}|$, 768), where each node has a 768-dimensional feature vector generated by CodeBERT. Edge connections are denoted by an index matrix of size (2, $|\mathcal{E}|$), with each column representing an edge's connected nodes. Since it is a directed graph, the first index of the column indicates the source node, and the second index represents the target node. Edge attributes are captured in a matrix sized ($|\mathcal{E}|$, 4), providing a 4-dimensional feature vector for each edge. Here, $|\mathcal{V}|$ and $|\mathcal{E}|$ denote the total number of nodes and edges in $\mathcal{G}$, respectively.

For commits with only a single hunk that results in a graph with one node and no edges ($|\mathcal{V}|=1$, $|\mathcal{E}|=0$), the edge index matrix and edge attribute matrix will be empty, and the node feature matrix will contain one 768-dimensional vector representing the single hunk's features.
To avoid the problem of the model tending to learn towards the majority class (non-vulnerability fixing commits), we up-sample the labeled graphs from the minority class until the graphs in both the training and validation sets are balanced across different classes. Only then do we begin the entire training process.

2) \textit{Graph learning.}
In this work, we choose GCNs~\cite{kipf2016gcn} as the component of basic GNN convolutional layers. 
To learn complex graph structures and edge attributes better, we introduce an edge attention mechanism in GCN. Each layer of the graph convolutional network performs the following operations: weighted adjacency matrix calculation, edge attention coefficient calculation and feature aggregation.
Specifically, for each edge type $r$, the model calculates the normalized adjacency matrix $\hat{A}_r$, then computes the attention coefficient $\alpha_{ij}^r$ for each edge from node $i$ to $j$ of type $r$, weighting the influence of node features. Finally, for each node $i$, aggregate the features of all neighboring nodes, weight and calculate to obtain new feature representation.




We set 2 GCN layers and use ReLU~\cite{glorot2011relu} in the middle to help the model better capture information from different types of nodes and edges in heterogeneous graphs. Then, we employ mean-pooling to aggregate the node features and obtain the graph representation.

3) \textit{Graph classification.}
Following the graph learning phase, we leverage a dropout layer~\cite{srivastava2014dropout} as a regularization method to mitigate overfitting during model training. To determine if a commit is fixing-related, we employ a fully connected (FC) layer. This layer transforms the learned graph feature representation into a 2-channel output. Each channel corresponds to the probability of the commit being a non-VFC or VFC, respectively. 

To address the class imbalance problem in VFC detection~\cite{vulfixminer}, we adopted a weighted loss function strategy. This approach assigns different weights to samples from different classes, making the model more attentive to the minority class (i.e., VFC). Specifically, the weighted cross-entropy loss function is defined as follows:
$L_w = -\frac{1}{N}\sum_{i=1}^N w_{y_i} [y_i \log(\hat{y}_i) + (1-y_i) \log(1-\hat{y}_i)]$,
where N is the total number of samples, $y_i$ is the true label of the $i$-th sample, $\hat{y}_i$ is the model's predicted probability for the $i$-th sample, and $w_{y_i}$ is the weight corresponding to the class of sample $i$.
We set weights $w_0$ and $w_1$, typically with $w_0 + w_1 = 2$, and inversely proportional to the sample ratio. For example, if the ratio of positive to negative samples is 1:9, we can set $w_1$ = 1.8 and $w_0$ = 0.2.

\section{Experimental Evaluation}
\subsection{Implementation}
Our implementation consists of approximately 2K lines of code in Python and Scala. 
To retrieve and preprocess pre-commit and post-commit code versions, we implement a Python script to analyze git diffs. This parser extracts relevant code changes and prepares them for further processing. To generate CPGs for both pre-commit and post-commit versions, we extend the Joern parser~\cite{Joern} using Scala scripts. 
The hunk correlation analysis, including caller-callee, data flow, and control dependencies extraction, was implemented using Python scripts. 
To detect code change pattern replication, we utilize the Levenshtein distance~\cite{yujian2007LSH} to implement the similarity computation.

The core of our CommitGNN model was built using PyTorch 1.11.0, leveraging its efficient tensor computations and deep learning capabilities. 
We extend this with PyTorch Geometric 2.5.3 to support our graph-based learning approach.
For code tokenization and embedding, we use the RobertaTokenizer~\cite{liu2019roberta} provided by the HuggingFace Transformers library, which is optimized for CodeBERT. Our weighted loss function is also implemented in PyTorch.

\subsection{Experimental Setup}

\textbf{Research Questions.} We design our evaluation to answer the following four research questions.

\begin{itemize}[leftmargin=1em]
\item \textbf{RQ4: Detection Effectiveness:} How is the effectiveness of \textsc{Fixseeker} in detecting VFCs, compared to current SOTA methods? (Sec.~\ref{sec:rq4})
\item \textbf{RQ5: Vulnerability Coverage:} How does \textsc{Fixseeker} perform in detecting different types of vulnerabilities? (Sec.~\ref{sec:rq5})
\item \textbf{RQ6: Feature Analysis:} What is the relative contribution of different correlation features to \textsc{Fixseeker}'s detection accuracy (Sec.~\ref{sec:rq6})
\item \textbf{RQ7: Computational Efficiency:} How efficiently does \textsc{Fixseeker} generate CommitHCGs, and how well does it scale across repositories of varying sizes and commit complexities? (Sec.~\ref{sec:rq7})
\end{itemize}

\noindent\textbf{Dataset.}
Because the dataset used by previous works only contained Java and Python projects, we built a new dataset containing vulnerabilities across four PLs: C/C++, Java, Python, and PHP to answer the RQs.  We additionally considered C/C++ and PHP because they contributed the largest portions of VFCs in our empirical study.
These languages are well-supported by the Joern parser, ensuring high-quality code property graph generation and analysis. 
In Section~\ref{sec:empirical_study}, we constructed the VFC dataset. 
We initially considered the other commits from the development history of 2,832 open-source projects in the VFC dataset as potential non-VFC candidates. However, among these commits, some contain security-related keywords in their messages as identified by Zhou et al.~\cite{zhou2017regex}, suggesting they might be silent VFCs that have not been explicitly documented. Since manually verifying the security nature of each suspicious commit would be impractical, we chose to exclude these commits from our non-VFC dataset to maintain dataset quality and avoid potential contamination from unconfirmed commits.

To evaluate our model from the perspectives of both balanced and imbalanced class classification, we randomly sampled from the non-VFC set to obtain a balanced dataset with a 1:1 ratio of VFC:non-VFC and an imbalanced dataset with a 1:25 ratio of VFC:non-VFC, 
the imbalanced ratio is following~\cite{estabrooks2004imbalance}. 
The non-VFC set is sampled by project, using the proportion of project non-VFCs in the total number of non-VFCs as the sampling ratio. 
The imbalanced dataset reflects the natural distribution in real-world scenarios. 
After constructing the dataset, we split it into training, validation, and test sets following the standard approach. We first divide the dataset into a training+validation set (80\%) and a test set (20\%). Then, we further split the training and validation sets at an 80\%/20\% ratio. We note that the training and validation sets undergo up-sampling using the Synthetic Minority Over-sampling Technique (SMOTE)~\cite{chawla2002smote} 
to reduce the imbalanced nature of the data. Specifically, SMOTE generates synthetic samples for the minority class (VFCs) by interpolating between existing VFC samples and their k-nearest neighbors rather than simply duplicating existing samples. 
The detailed number of the datasets is shown in Table~\ref{tab:dataset}.

    

\begin{table}[h]\small
    \centering
    \caption{The Distribution of Balanced and Imbalanced Dataset}
    \vspace{-2mm}
    \begin{tabular}{l|llll|llll}
        \Xhline{0.8pt}
       \multirow{2}*{}  & \multicolumn{4}{c}{\bf Balanced} & \multicolumn{4}{c}{\bf Imbalanced} \\
       \cline{2-9}
       & C/C++ & Java & Python &  PHP & C/C++ & Java & Python &  PHP \\
       \hline\hline
       {\bf \#VFC}  & 5,168 & 768 & 1,147 & 3,157 & 5,168 & 768 & 1,147 & 3,157\\
       \hline
      {\bf \#non-VFC}  &  5,168 & 768 & 1,147 & 3,157 & 129,200 & 19,200 & 28,675 & 78,925\\
      
       \Xhline{0.8pt}
    \end{tabular}
    
    \label{tab:dataset}
\end{table}

\noindent\textbf{Metrics.}
In the balanced dataset, we use three metrics: $Precision=\frac{\left |identified_{v}\cap correct_{v} \right |}{\left |identified_{v}\right |} $, 
$Recall=\frac{\left |identified_{v}\cap correct_{v}\right |}{\left |correct_{v}\right|}$
and $F1=\frac{2\times Precision \times Recall}{Precision + Recall}$,
where $|identified_{v} \cap correct_{v}|$ represents the number of VFCs correctly identified by the model, $|identified_{v}|$ represents the total number of commits identified as VFCs by the model and $|correct_{v}|$ represents the total number of actual VFCs.

In the imbalanced dataset, in addition to the above three metrics, we use three other metrics: AUC-ROC, AUC-PR, and effort-aware metrics CostEffort@L to measure the effectiveness of VFC detection, which are used by SOTA works~\cite{vulfixminer, nguyen2023midas, wang2021patchrnn}.

\textit{AUC-ROC} is an area under the Receiver Operating Characteristic (ROC) curve~\cite{hanley1982ROC}. It is a metric for binary classification models, representing the ability to distinguish between classes across all possible thresholds. The score is mathematically defined as:
$\text{AUC-ROC} = \frac{\sum_{i \in \text{positiveClass}} \text{rank}_i - \frac{M(1+M)}{2}}{M \times N}$,
where $\text{rank}_i$ is the rank of the $i$-th positive sample in the ranked list of predictions, $M$ is the number of positive examples, and $N$ is the number of negative samples. 
AUC-ROC values range from 0 to 1, with 0.5 representing random guessing and 1 indicating perfect classification. 
%
AUC benefits from its insensitivity to class imbalance and its ability to provide a comprehensive performance measure without requiring a specific classification threshold. 

\textit{AUC-PR}~\cite{davis2006relationship} is an area under the Precision-Recall (PR) curve. It is a metric for binary classification models, particularly useful for imbalanced datasets, representing the trade-off between precision and recall across various classification thresholds. The AUC-PR score is mathematically defined as:
$\text{AUC-PR} = \int_{0}^{1} P(R) \, dR$,
where $P$ is precision, $R$ is recall, and the integral is computed over all possible recall values. AUC-PR values range from 0 to 1, with higher values indicating better performance. Unlike AUC-ROC, AUC-PR focuses on the performance of the positive class and is not affected by the large number of true negatives, making it more informative in identifying positive instances accurately.

\textit{CostEffort@L} measures the proportion of VFCs that the model can detect when inspecting L\% of the total lines of code. First, commits are ranked from high to low based on the probabilities predicted by the model, then the number of actual VFCs within the top L\% of code lines is counted. This metric can be represented as: (number of VFCs detected in the top L\% of code lines) / (total number of VFCs). The higher the value of CostEffort@L, the better the model's performance. We considered CostEffort under multiple L values (i.e., 5\%and 20\%) to evaluate the model's performance under different workload constraints.

\vspace{4px}
\noindent\textbf{Baseline.}
We compare our method with the following three baselines:

1) PatchRNN~\cite{wang2021patchrnn} is a deep learning system for identifying security patches, using both commit messages and diff code as features, employing TextRNN~\cite{lai2015textrnn} for commit messages and a twin RNN for diff code. Because our task scenario only considers the code changes in commits, we remove the commit messages feature extraction and training from PatchRNN.

2) VFFINDER~\cite{nguyen2023vffinder} is a graph-based approach that captures code structural changes using annotated Abstract Syntax Trees (ASTs) for silent vulnerability fix identification. It utilizes a graph attention network (GAT) to extract features from ASTs and distinguish VFC from non-VFC. However, VFFINDER is currently limited to C/C++ projects.

3) VulFixMiner~\cite{vulfixminer} is a Transformer-based model for detecting silent vulnerability fixes in code commits. It uses CodeBERT~\cite{feng2020codebert} to extract semantic features from file-level code changes, then aggregates them to the commit level.

4) Midas~\cite{nguyen2023midas} is a multi-granularity deep learning model for detecting VFCs. It extracts features at commit, file, hunk, and line levels, using CodeBERT for code change representation and various neural networks for feature extraction. A neural classifier makes the final prediction.

\subsection{Detection Effectiveness (RQ4)}
\label{sec:rq4}
\subsubsection{Overall Results} 
As illustrated in Table~\ref{tab:comparison}, the \textsc{Fixseeker} system demonstrates strong performance across multiple PLs. On the balanced datasets, \textsc{Fixseeker} achieves the highest F1-score for C/C++ (0.8577), Java (0.8447), Python (0.8319), and PHP (0.8272). 
\textsc{Fixseeker} maintains robust performance on the imbalanced datasets with F1 scores of 0.4991, 0.4191, 0.4896, and 0.4322 for C/C++, Java, Python, and PHP respectively. These F1 scores, though lower than those on balanced datasets due to the inherent challenge of class imbalance, still outperform baseline approaches.
Additionally, \textsc{Fixseeker} consistently outperforms other methods in terms of AUC-ROC and AUC-PR metrics. For instance, in the Python dataset, \textsc{Fixseeker} achieves an AUC-ROC of 0.8344 and an AUC-PR of 0.7856. 
It is important to note that the performances on balanced and imbalanced datasets are not directly comparable due to their different data distributions, reflecting various real-world scenarios of vulnerability detection tasks.

\subsubsection{Comparison with RNN and Graph-based Approaches}
We compare Fixseeker with PatchRNN and VFFINDER over our ground-truth dataset by applying the same training and test set splitting. Because VFFINDER is limited to C/C++ projects, we only compare the results of VFFINDER and Fixseeker in the C/C++ dataset in Table~\ref{tab:comparison}.
\textsc{Fixseeker} significantly outperforms PatchRNN with 9.0\% higher F1 in the balanced dataset and 22.6\% higher F1, 12.1\% higher AUC-ROC and 38.8\% higher AUC-PR in the imbalanced dataset on average. The main advantage of \textsc{Fixseeker} over PatchRNN stems from its ability to capture complex hunk relationships through graph structures. While PatchRNN processes code changes as sequential data, it struggles to model the intricate dependencies between different hunks, such as caller-callee relationships and data flow dependencies that are naturally represented in our graph structure. 

Although both \textsc{Fixseeker} and VFFINDER utilize graph neural networks, \textsc{Fixseeker} demonstrates superior performance with 6.6\% and 21.26\% higher F1-score in balanced and imbalanced datasets. This improvement can be attributed to two key factors: (1) Feature extraction: While VFFINDER focuses on structural changes at the AST level, Fixseeker extracts richer features by analyzing the relationships between different hunks, capturing both syntactic and semantic dependencies that are crucial for vulnerability fix identification. (2) Network architecture: Our edge-aware attention mechanism in GCNs shows better performance than VFFINDER's GAT architecture in capturing hunk-level correlations. This design allows the model to learn different attention weights for different types of hunk relationships during feature aggregation. In contrast, VFFINDER's GAT treats all structural relationships uniformly when computing attention.

\subsubsection{Comparison with SOTA VFC Detection Approaches} 
We compare \textsc{Fixseeker} with baseline works over datasets for C/C++, Java, Python and PHP by applying the same training and test set splitting. 
The experimental results are summarized in Table~\ref{tab:comparison}. 
On the balanced dataset, \textsc{Fixseeker} outperforms other approaches in most metrics across different PLs. For C/C++ and Python datasets, \textsc{Fixseeker} achieves the highest F1 and precision, though Midas has the highest recall at 0.9039. On the Java dataset, \textsc{Fixseeker} excels in recall and F1, but VulFixMiner leads in precision with 0.8095.
On the imbalanced dataset, \textsc{Fixseeker} demonstrates consistently superior performance across all evaluation metrics. Specifically, \textsc{Fixseeker} improves the F1-score by 7.1\% to 59.7\% compared to the best baseline methods. For example, on the C/C++ dataset, \textsc{Fixseeker} achieves an F1 of 0.4991, outperforming Midas (0.2891), VulFixMiner (0.3125), and PatchRNN (0.2347).
While Midas shows competitive performance in AUC-ROC, \textsc{Fixseeker} maintains a notable advantage in AUC-PR scores, improving by 8.24\% on average.

Due to the extreme imbalance between non-VFC and VFC in OSS (VFCs typically account for a small percentage), false positives matter more than false negatives. Therefore, our method aims to reduce the false positive rate while maintaining a high recall. The consistently high AUC-PR scores of \textsc{Fixseeker} across languages (e.g., 0.8020 for C/C++, 0.7985 for Java) demonstrate its effectiveness in handling this imbalance, outperforming other approaches in most cases.

\begin{table}[tp]\footnotesize
    \centering
    \caption{Performance of \textsc{Fixseeker} and Baseline Works on Different Program Language Projects}
    \vspace{-2mm}
    \begin{tabular}{c|c|c c c|c c c c c}
        \Xhline{0.8pt}
        \multirow{3}*{\textbf{PL}}& \multirow{3}*{\textbf{Method}} & \multicolumn{3}{c|}{\textbf{Balanced Dataset}} & \multicolumn{5}{c}{\textbf{Imbalanced Dataset}} \\
        \cline{3-10}
        ~ & ~ & \multirow{2}*{Precision} & \multirow{2}*{Recall} & \multirow{2}*{F1}  & \multirow{2}*{F1} & \multirow{2}*{AUC-ROC} & \multirow{2}*{AUC-PR} & \multicolumn{2}{c}{CostEffort} \\
        \cline{9-10}
        ~ & ~ & ~ & ~ & ~  & ~ & ~& ~ & @5 & @20 \\
        \hline\hline
        \multirow{5}*{C/C++} & PatchRNN & 0.7391 & 0.8143 & 0.7749 & 0.2347 &0.6852 & 0.4201 & 0.3259 & 0.7467\\
        ~ & VFFINDER & 0.7843 & 0.8291 & 0.7917 & 0.2865 & 0.7538 & 0.5104 & 0.3973 & 0.7896\\
        ~ & VulFixMiner & 0.8045 & 0.7692 & 0.7865 & 0.3125 & 0.7544 & 0.6321 & 0.3813 & 0.8210\\
        ~ & Midas & 0.7798 & \cellcolor{lightgray}0.9039 & 0.8372 & 0.2891 &0.8015 & 0.7611 & 0.5836 & 0.8812\\
        ~ & \textsc{Fixseeker} & \cellcolor{lightgray}0.8178 & 0.9016 & \cellcolor{lightgray}0.8577 & \cellcolor{lightgray}0.4991 & \cellcolor{lightgray}0.8075 & 0.8020\cellcolor{lightgray} & \cellcolor{lightgray}0.6429 & \cellcolor{lightgray}0.9000 \\
        \hline
        \multirow{4}*{Java} & PatchRNN & 0.7435 & 0.8202 & 0.7800 & 0.2297 &0.7539 & 0.4615 & 0.3548 & 0.5161\\
        ~ & VulFixMiner & \cellcolor{lightgray}0.8095 & 0.7727 & 0.7907 &0.3590& 0.8088 & 0.6593 & 0.3333 & 0.7778\\
        ~ & Midas & 0.7966  & 0.8943 & 0.8426 & 0.3915 & 0.8324 & 0.7187 & 0.7016 & 0.9120\\
        ~ & \textsc{Fixseeker} & 0.7768 & \cellcolor{lightgray}0.9255 & \cellcolor{lightgray}0.8447 & \cellcolor{lightgray}0.4191 & \cellcolor{lightgray}0.8357 & \cellcolor{lightgray}0.7985 & \cellcolor{lightgray}0.8078 & \cellcolor{lightgray}0.9374\\
        \hline
        \multirow{4}*{Python} & PatchRNN & 0.6874 & 0.7947 & 0.7372 & 0.2564& 0.6985 & 0.2681 & 0.2438 & 0.5655\\
        ~ & VulFixMiner & 0.6511 & 0.8402 & 0.7337 & 0.3077 & 0.8095 & 0.6932 & 0.3636 & 0.7272\\
        ~ & Midas & 0.7458 & \cellcolor{lightgray}0.9333 & 0.8291 &0.3846 &0.8073 & 0.7259 & 0.6516 & 0.8502\\
        ~ & \textsc{Fixseeker} & \cellcolor{lightgray}0.7684 & 0.9068 & \cellcolor{lightgray}0.8319 & \cellcolor{lightgray}0.4896 & \cellcolor{lightgray}0.8344 & \cellcolor{lightgray}0.7856 & \cellcolor{lightgray}0.7545 & \cellcolor{lightgray}0.9281\\
        \hline
        \multirow{4}*{PHP} & PatchRNN & 0.6523 & 0.7812 & 0.7108 & 0.2159 & 0.6791 & 0.4543 & 0.3312 & 0.6421 \\
        ~ & VulFixMiner & 0.6892 & 0.8156 & 0.7468 & 0.3417 & 0.7985 & 0.6754 & 0.3512 & 0.7103 \\
        ~ & Midas & 0.7312 & 0.8945 & 0.8046 & 0.3528 & 0.8086 & 0.7102 & \cellcolor{lightgray}0.6534 & 0.8667 \\
        ~ & \textsc{Fixseeker} & \cellcolor{lightgray}0.7578 & \cellcolor{lightgray}0.9106 & \cellcolor{lightgray}0.8272 & \cellcolor{lightgray}0.4322 & \cellcolor{lightgray}0.8223 & \cellcolor{lightgray}0.7689 & 0.6321 & \cellcolor{lightgray}0.9104 \\
        \Xhline{0.8pt}
    \end{tabular}
    \label{tab:comparison}
\end{table}

\subsubsection{Effort-aware Performance} 
In real-world scenarios, it's crucial to detect VFCs efficiently with minimal inspection effort. To evaluate this aspect, we use CostEffort@5 and CostEffort@20 metrics, which measure the percentage of VFCs detected when inspecting 5\% and 20\% of the total lines of code, respectively. When performing \textsc{Fixseeker} on the imbalanced datasets, Table~\ref{tab:comparison} shows high CostEffort@5 and CostEffort@20 scores across languages, indicating that \textsc{Fixseeker} can detect a high percentage of VFCs with minimal inspection effort. For example, in C/C++, VULSEEKER achieves a CostEffort@5 of 0.6429, outperforming VulFixMiner (0.3813) and Midas (0.5836). 

\subsubsection{Case Studies}
To learn how \textsc{Fixseeker} outperforms other approaches, we study VFC cases only detected by \textsc{Fixseeker} and exemplify the scenarios as follows.

\begin{figure}
    \begin{subfigure}{0.49\textwidth}
    \includegraphics[width=0.96\linewidth]{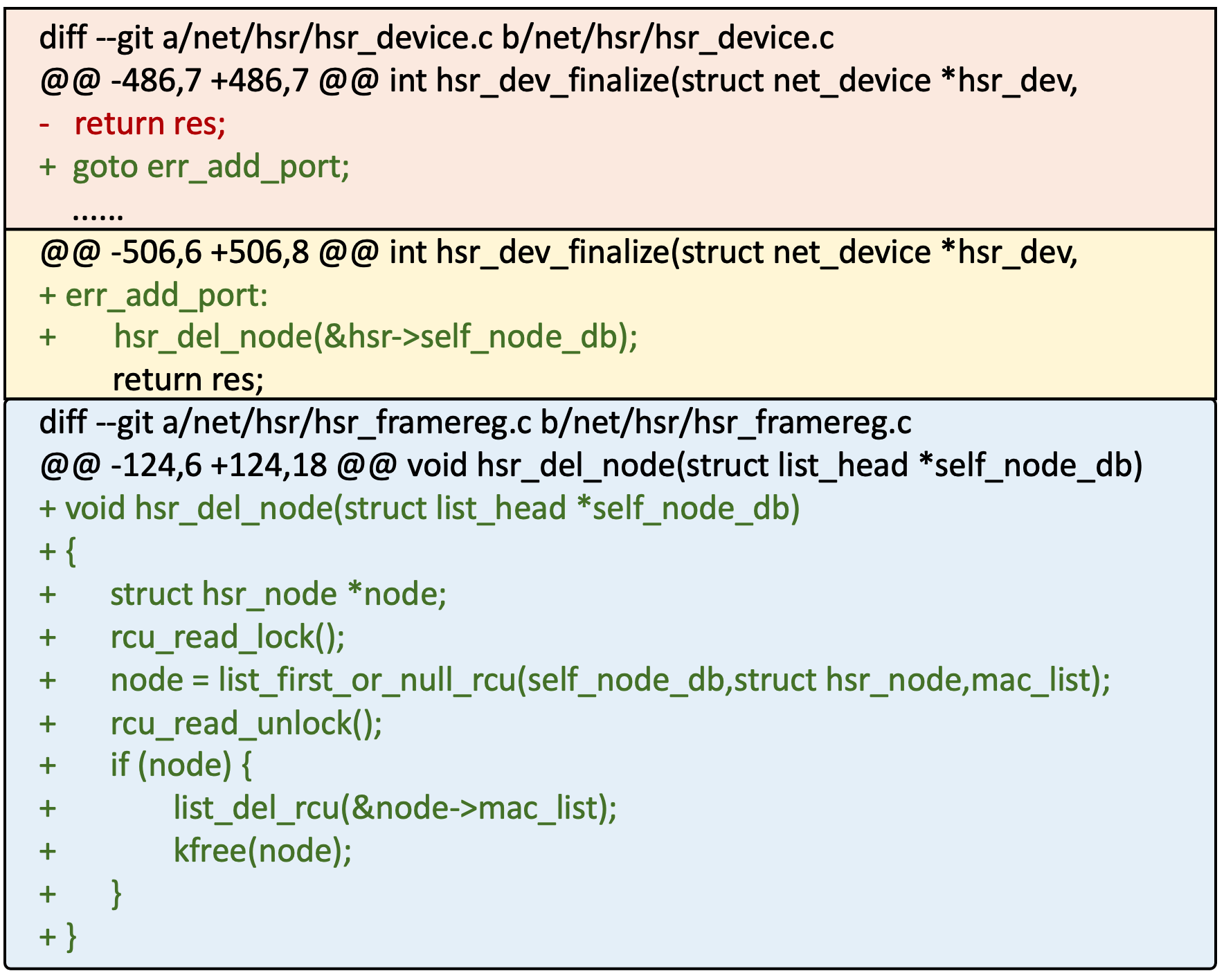}
    \caption{VFC of CVE-2019-16995} 
    \label{fig:caseA}
    \end{subfigure}
    \hfill
    \begin{subfigure}{0.49\textwidth}
    \includegraphics[width=0.96\linewidth]{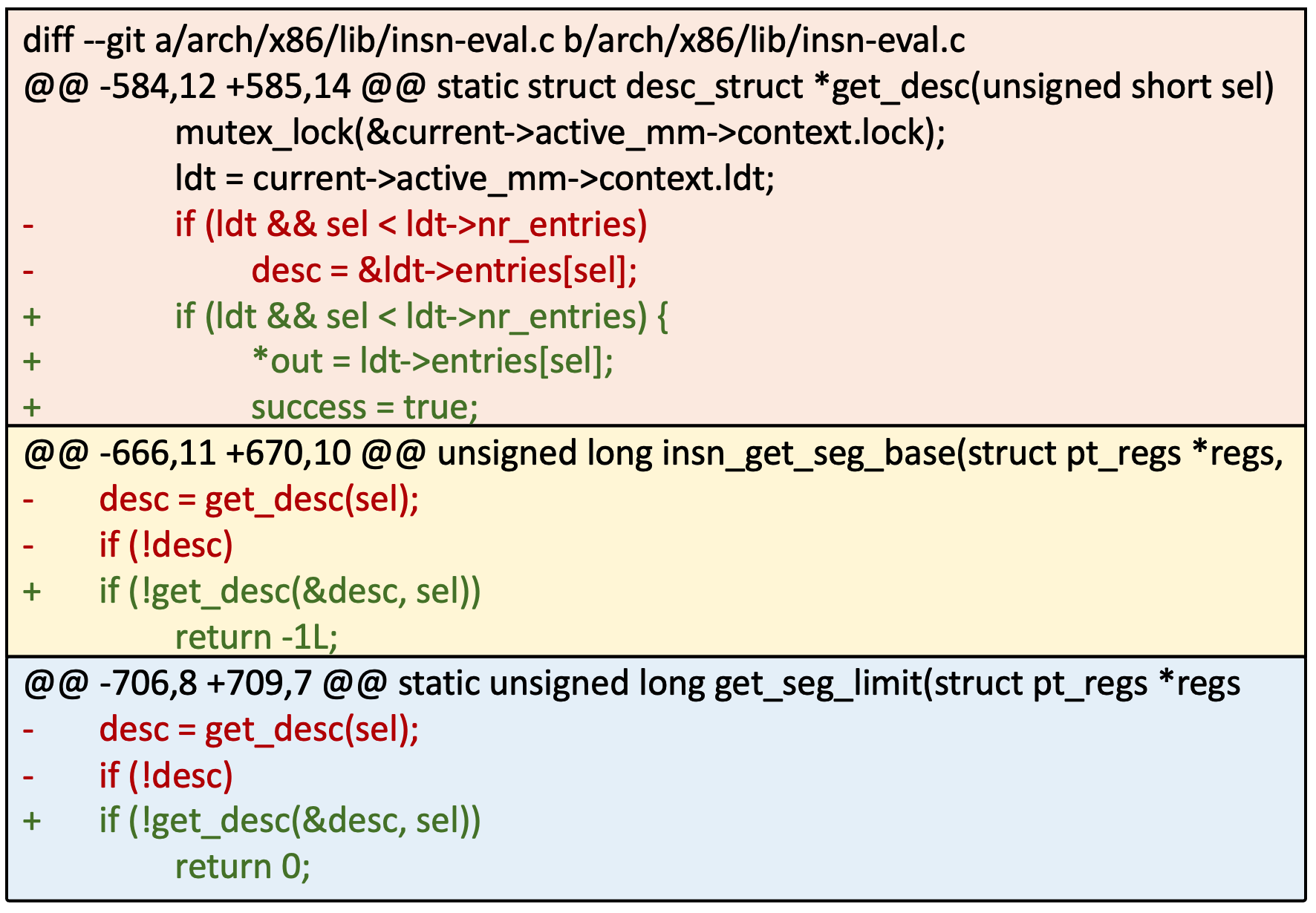}
    \caption{VFC of CVE-2019-13233}
    \label{fig:caseB}
    \end{subfigure}

\caption{The Cases only Detected by \textsc{Fixseeker}}
    \label{fig:3mthods}
    \vspace{-3mm}
\end{figure}

1) The fixes are spread across different files.
\textsc{Fixseeker} specifically analyzes the correlations between hunks, making it more adept at detecting related fixes across multiple files.
The commit in Figure~\ref{fig:caseA} shows changes in two files \texttt{hsr\_device.c} and \texttt{hsr\_framereg.c}. In \texttt{hsr\_device.c}, there's a modification in the error handling logic, replacing a simple \texttt{return res;} with \texttt{goto err\_add\_port;}. This change is correlated with the addition of the \texttt{hsr\_del\_node()} function call. The \texttt{hsr\_del\_node()} function is then implemented in \texttt{hsr\_framereg.c}, where it's defined with proper error handling and memory management. \textsc{Fixseeker}'s ability to analyze relationships between these hunks across different files allows it to recognize this as a coordinated fix for a potential resource leak or error-handling vulnerability, which might be missed by approaches that only consider adjacent hunks or line-level change features.

2) Vulnerabilities that require synchronized changes across multiple functions. For example, simultaneously modifying the definition of an interface function and multiple places where this function is called. 
The changes shown in Figure~\ref{fig:caseB} involve the modification of the \texttt{get\_desc} function and its usage in two other functions.
In the \texttt{get\_desc} function, the logic is altered to use a pointer \texttt{out} and a \texttt{success} flag, potentially improving error handling. Correspondingly, the calls to \texttt{get\_desc} in \texttt{insn\_get\_seg\_base} and \texttt{get\_seg\_limit} functions are modified to adapt to this new interface. 
The new version passes \texttt{\&desc} as an argument and checks the return value of the function.
This coordinated change across multiple functions demonstrates a consistent modification pattern that addresses a potential vulnerability in how descriptor information is retrieved and handled. VulSeeker's ability to analyze similar code change patterns between different hunks allows it to recognize this as a coordinated fix, which is overlooked by SOTA approaches.

\subsection{Vulnerability Coverage (RQ5)}
\label{sec:rq5}
As a large-scale real-world dataset, our dataset allows us to evaluate system performance across different vulnerability types. Table~\ref{tab:type_performance} illustrates \textsc{Fixseeker}'s performance for various vulnerability types, ranked by CWE ID. We also list the proportion of the Top 10 popular vulnerability types and the corresponding recall of \textsc{Fixseeker}. Recall refers to the percentage of correctly detected samples among all VFCs.
Through analyzing \textsc{Fixseeker}'s performance on each vulnerability type, we draw two important conclusions:

1) \textsc{Fixseeker} excels in detecting certain vulnerability types. 
This variation reflects the uniqueness of code patterns in different types of vulnerability fixes. For example:
Memory-related vulnerabilities (i.e., CWE-476, CWE-119, CWE-125 and CWE-787) generally have high detection rates, with recalls exceeding 0.75. This might be due to these vulnerabilities often involving explicit memory operations and boundary check patterns that our model can easily recognize.
Integer overflow (CWE-190) and buffer copy (CWE-120) issues, despite having fewer samples, still show good detection results with recalls of 0.7221 and 0.7903, respectively. This indicates our model's ability to generalize well to vulnerabilities with specific patterns.


2) Some vulnerability types show relatively lower detection performance, reflecting the complexity of their fix patterns. Specifically: Cross-site scripting (CWE-79) and improper input validation (CWE-20) have relatively lower recalls of 0.6492 and 0.6200, respectively. This may be because fixes for these vulnerabilities often involve diverse context handling and complex validation logic, making pattern recognition more difficult. For example, a cross-site scripting fix might require understanding both the input sanitization code and how the sanitized data is used across different functions and files, while an input validation fix often involves multiple validation checks distributed across different program paths. Future work could explore incorporating program analysis techniques that can better track such cross-function and cross-file dependencies.


\begin{table}[h]\small
    \centering
    \caption{Performance of \textsc{Fixseeker} over Different Vulnerability Types}
    \vspace{-2mm}
    \begin{tabular}{l|c|c|c}
        \Xhline{0.8pt}
         \textbf{CWE ID} & \textbf{Description} & \textbf{Percentage} & \textbf{Recall}  \\
         \hline\hline
         CWE-79 & Cross Site Scripting & 16\% & 0.6492\\
         CWE-125 & Out-of-bounds Read & 4\% & 0.7851 \\
         CWE-787 & Out-of-bounds Write & 3\%& 0.7623 \\   
         CWE-119 & Improper Restriction within Memory Buffer Bounds& 3\% & 0.8286 \\
         CWE-20 & Improper Input Validation& 3\% & 0.6200 \\
         CWE-476 & Null Pointer Dereference & 3\% & 0.8430 \\
         CWE-190 & Integer Overflow or Wraparound  & 2\% & 0.7221 \\
         CWE-120 & Buffer Copy without Checking Size of Input & 1\% & 0.7903\\
        \Xhline{0.8pt}
    \end{tabular}
    \label{tab:type_performance}
\end{table}

\subsection{Feature Analysis (RQ6)}
\label{sec:rq6}
In this section, we explore the impact of four specific edge (correlation) features outlined in Table~\ref{tab:correlation} on the performance of \textsc{Fixseeker}.
We utilized the GNNExplainer~\cite{ying2019gnnexplainer} masking generation algorithm in the imbalanced dataset to generate corresponding edge masks for the input instance graphs, identifying the edge features that have the most significant impact on prediction. Figure x illustrates the importance of edge features for different batches in the test set, which is measured by $edge\_weight$ in GNNExplainer, with a range from 0 to 1.
All features show median importance values around 0.50, indicating significant contributions to model predictions. The Caller-Callee Dependency feature demonstrates the widest range and highest peak, suggesting this correlation is often crucial in identifying vulnerability-fixing commits. Data using and Control Dependency features exhibit similar shapes, indicating comparable importance. The Similar Pattern feature has the narrowest distribution, suggesting moderate but consistent significance.

\begin{figure}[h]
    \vspace{-3mm}
    \centering
    \includegraphics[width=0.55\linewidth]{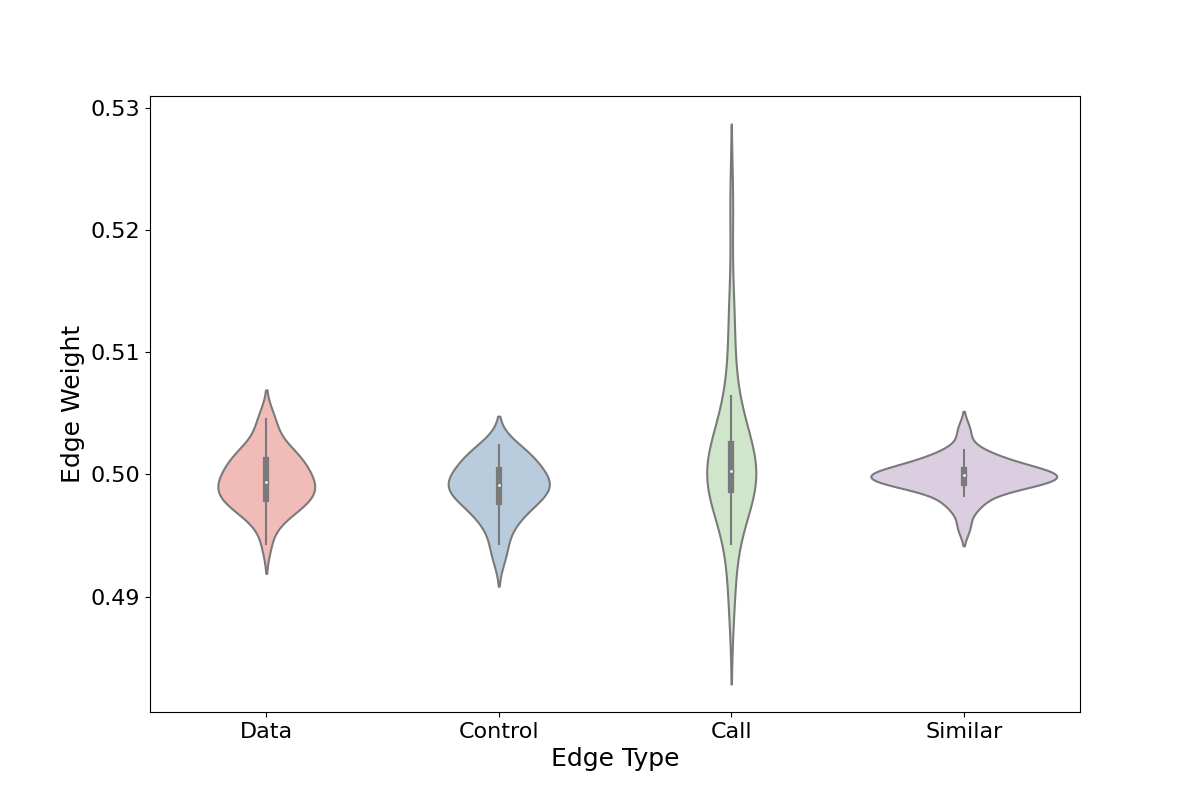}
    \caption{The Weight of Different Correlation Features in Imbalanced Dataset}
    \label{fig:voilin}
\end{figure}
\vspace{-2mm}

\subsection{Computational Efficiency (RQ7)}
\label{sec:rq7}
As illustrated in Table~\ref{tab:overhead}, we evaluated the overhead of CommitHCG generation across six repositories. The size of these repositories ranges from 34MB to 784MB, with the number of commits varying from 80 to 179. The number of hunks per repository spans from 337 to 1474.
The average number of nodes in the generated CommitHCGs ranges from 2.79 to 18.43, while the average number of edges varies from 2.21 to 28.62. This suggests that the complexity of the generated graphs is manageable and appropriate for our model.
The main overhead of our approach is in constructing CommitHCGs for given commits. The last column shows that the average time to generate a CommitHCG ranges from 12.06s to 30.92s across different repositories. Notably, the generation time does not directly correlate with repository size or number of commits but appears more related to the complexity of the changes (as indicated by the number of hunks and average nodes/edges).
These results demonstrate that our CommitHCG generation process is efficient and scalable across various repository sizes and commit complexities, with generation times that are acceptable for practical use in vulnerability detection workflows.
\begin{table}[h]\small
    \centering
    \caption{Overhead of CommitHCG generation}
    \vspace{-2mm}
    \begin{tabular}{c|c|c|c|c|c|c}
        \Xhline{0.8pt}
        \textbf{Repo} & \textbf{Size} & \textbf{\#Commit} & \textbf{\#Hunk} & \textbf{\#Avg. Node} & \textbf{\#Avg. Edge} & \textbf{\#Avg. Time} \\
        \hline\hline
        ImageMagick & 213M & 179 & 699 & 3.91 & 5.22 & 24.39s\\  
        Gpac & 199MB & 137 & 533 & 3.89 & 4.72 & 16.63s\\ 
        FFmpeg & 516MB & 121 & 337 & 2.79 & 2.21 & 16.92s\\
        Tcpdump & 34MB & 117 & 1014 & 8.67 & 28.62 & 23.06s\\
        Cpython & 646MB & 87 & 890 & 9.18 & 5.52 & 19.67s\\
        Xwiki & 784MB & 80 & 1474 & 18.43 & 8.69 & 20.58s\\
        \Xhline{0.8pt}
    \end{tabular}
    \label{tab:overhead}
\end{table}

\section{Discussion}
\textbf{Practicality.} 
Due to the prevalence of silent vulnerability fixes in open-source software, traditional methods relying on commit messages or explicit security labeling may fail to identify critical security patches. Fixseeker addresses this challenge by focusing solely on code diffs and leveraging advanced hunk correlation analysis. 
This approach allows Fixseeker to outperform existing methods, particularly in detecting complex, multi-file vulnerability fixes that may not be apparent through surface-level analysis. 
\textsc{Fixseeker}'s language-agnostic nature, focusing on structural and semantic code changes rather than language-specific features, makes it applicable across various programming languages. This versatility allows security teams and developers to maintain consistent security practices across diverse technology stacks.
Additionally, Fixseeker can prioritize code reviews for potentially security-relevant commits, reducing vulnerability windows. 

\vspace{4px}
\noindent\textbf{Limitation and Future Work.}
As we mentioned in the dataset description in Section 5, we used regular expressions to filter out many suspicious commits, as we could not definitively determine whether they were vulnerability-fixing commits (VFCs). \textsc{Fixseeker}, while necessary for creating a reliable dataset, potentially excludes a significant number of actual VFCs. In future work, we plan to leverage manual verification to confirm whether these suspicious commits are vulnerability-related. This effort will help expand and enrich our dataset, potentially improving the model's ability to detect a wider range of vulnerability-fixing patterns.

A significant portion of \textsc{Fixseeker}'s processing time is spent on generating and analyzing CPGs using the static analysis tool, Joern parser. While this step is crucial for extracting rich semantic information from code changes, it can be time-consuming, especially for large commits. 
Our future work will focus on optimizing this process, possibly by exploring parallelizing CPG generation or developing more efficient static analysis methods tailored to vulnerability detection, ensuring \textsc{Fixseeker} can operate effectively in real-time within fast-paced development workflows.

\section{Related Work}
\textbf{Empirical Studies on Vulnerability Fixing.}
To mine the hunks correlations in VFCs, we conducted an empirical study on a large-scale dataset. 
Many research studies~\cite{xu2022tracking,antal2020exploring,li2017large,liu2020large,iannone2022secret,chinthanet2021lags,tan2022understanding} have investigated vulnerability fixing (security patches) from other perspectives.
Xu et al.~\cite{xu2022tracking} conducted an empirical study to understand the quality and characteristics of patches for OSS vulnerabilities in two industrial vulnerability databases.
Li et al.~\cite{li2017large} 
found that security fixes are smaller than non-security ones, vulnerabilities often exist for years, and a significant percentage of patches are flawed or incomplete.
Emanuele et al.~\cite{iannone2022secret} 
examined vulnerabilities' introduction and removal and found that vulnerabilities often result from multiple commits and remain unfixed for over a year. 
Chinthanet et al.~\cite{chinthanet2021lags} examines delays in adopting security fixes in npm packages. It found that fixing releases often include unrelated changes. 
Tan et al.~\cite{tan2022understanding} 
found that 80\% of CVE-Branch pairs are unpatched, posing significant risks. Patch porting is time-consuming, averaging 40.46 days.
Some studies~\cite{wang2020empirical, wang2019detecting} aim to investigate the secret security patch in open-source software.
Wang et al.~\cite{wang2020empirical} 
developed a machine learning tool to distinguish security from non-security patches, analyzing three SSL libraries to highlight the need for better patch management.
Wang et al.~\cite{wang2019detecting} 
analyzed over 4,700 known security patches and discovered 12 secret patches in SSL libraries.
Different from these, our work is the first to systematically study the correlations between different code hunks within vulnerability fixes. Our empirical findings 
motivate the design of \textsc{Fixseeker}.

\vspace{4px}
\noindent\textbf{VFC Detection.}
Due to \textsc{Fixseeker}'s focus on detecting silent vulnerability fixes, we only consider code diffs. The works most relevant to \textsc{Fixseeker}'s task objectives are VulFixMiner and MiDas. 
VulFixMiner~\cite{vulfixminer} uses a Transformer-based model, which automatically extracts semantic meaning from commit-level code changes to identify silent vulnerability fixes in open-source software.
Midas~\cite{nguyen2023midas} proposed a multi-granularity model for detecting the VFC in open-source software. It analyzes code changes at different levels and uses an ensemble approach, improving detection accuracy and efficiency.
Many other approaches are proposed to detect VFCs with the help of information in commit messages and security issues.
Sabetta et al.~\cite{sabetta2018practical} treat code changes as natural language documents, using document classification methods to identify security-relevant commits in open-source software repositories.
Hermes~\cite{nguyen2022hermes} proposed a commit-issue link recovery technique to infer the potential missing link, incorporating the information from issue trackers to boost the VFC classifier.
Vulcurator~\cite{nguyen2022vulcurator} is a preliminary work that combines commit messages, code changes and issue reports using deep learning for VFC classification. As its code change analysis component later evolved into VulFixMiner~\cite{vulfixminer}, we chose to compare our approach with the more mature VulFixMiner in our evaluation.
VFCFinder~\cite{dunlap2024vfcfinder} used NL-PL models to generate the top-five ranked set of VFCs for a given security advisory.
Unlike previous works that rely heavily on commit messages and issue reports, or treat code changes as flat sequences or isolated units, our work is the first to model the complex relationships between different code hunks using a graph-based approach. By capturing these inter-hunk correlations through our CommitHCG representation, we can better understand the scope and impact of vulnerability fixes, improving detection accuracy.

\vspace{4px}
\noindent\textbf{Vulnerability-related Task with Graph Representation.}
\textsc{Fixseeker} employs a GNN model to learn from a graph where hunks serve as nodes and correlations as edges, aiding in VFC classification. Many other methods~\cite{hu2023interpreters,hin2022linevd,qiu2024vulnerability,chu2024graph,nguyen2022regvd} also use graph learning to address vulnerability-related tasks.
LineVD~\cite{hin2022linevd} combined graph neural networks and transformers to analyze code dependencies and tokens for statement-level vulnerability detection in software.
Qiu et al.~\cite{qiu2024vulnerability} introduced a heterogeneous GNN framework for predicting specific vulnerability types in code. 
Chu et al.~\cite{chu2024graph} proposed a novel counterfactual explainer for GNN-based vulnerability detection. It identifies minimal code graph changes that alter predictions, helping to understand and fix vulnerabilities.
ReGVD~\cite{nguyen2022regvd} is a language-independent GNN model for vulnerability detection. It combines token sequences with pre-trained language models to analyze code vulnerabilities.

\section{Conclusion}
We have conducted a large-scale empirical study to understand the characteristics of VFCs across six popular programming languages, revealing the prevalence and importance of multi-hunk correlations in VFCs. 
We have proposed \textsc{Fixseeker}, a graph-based approach to detect silent vulnerability fixes in open-source software. Our comprehensive evaluation has demonstrated the effectiveness, efficiency, and generality of \textsc{Fixseeker} across multiple programming languages and various vulnerability types. We have shown that \textsc{Fixseeker} outperforms state-of-the-art approaches on different metrics.

\section{Data Availability}
The data and the implementation of \textsc{Fixseeker} are publicly available at \url{https://github.com/Veronica-L/fixseeker}.
\bibliographystyle{ACM-Reference-Format}
\bibliography{main}

\appendix

\end{document}